\documentclass[final, 5p, twocolumn, 11pt]{elsarticle}

\usepackage{graphics}
\usepackage{amssymb}
\usepackage{amsmath}
\usepackage[hidelinks]{hyperref}
\usepackage{lineno}
\usepackage{xcolor}

\definecolor{skyblue}{HTML}{005398}
\hypersetup{colorlinks=true, allcolors=skyblue}

\newcommand{\legolas}{\textsf{Legolas}}
\newcommand{\pylbo}{\textsf{Pylbo}}
\newcommand{\legolasOne}{\textsf{Legolas 1.x}}
\newcommand{\legolasTwo}{\textsf{Legolas 2.0}}
\newcommand{\reg}{\textsuperscript{\textregistered}}

\newcommand\T{\rule{0pt}{3ex}}			

\newcommand{\amat}{\boldsymbol{A}}
\newcommand{\bmat}{\boldsymbol{B}}
\newcommand{\Lmat}{\boldsymbol{L}}
\newcommand{\Umat}{\boldsymbol{U}}
\newcommand{\Gmat}{\boldsymbol{G}}
\newcommand{\uvec}{\boldsymbol{u}}
\newcommand{\Rvec}{\boldsymbol{R}}
\newcommand{\hermconj}{\mathsf{H}}
\newcommand{\bfx}{\boldsymbol{x}}
\newcommand{\wreal}{\omega_\text{R}}
\newcommand{\wimag}{\omega_\text{I}}
\newcommand{\bfv}{\boldsymbol{v}}
\newcommand{\bfa}{\boldsymbol{a}}
\newcommand{\bfb}{\boldsymbol{B}}
\newcommand{\itxt}{\text{i}}
\newcommand{\ipone}{{\text{i} + 1}}
\newcommand{\unit}[1]{\boldsymbol{e_#1}}

\journal{Computer Physics Communications}

\begin{document}
\begin{frontmatter}

\title{Legolas 2.0: Improvements and extensions to an MHD spectroscopic framework}

\author[a]{Niels Claes}\corref{author}
\author[a]{Rony Keppens}
\cortext[author] {Corresponding author.\\\textit{E-mail address:} niels.claes@kuleuven.be}
\address[a]{Centre for mathematical Plasma-Astrophysics, KU Leuven, Celestijnenlaan 200B, 3001 Leuven, Belgium}


\begin{abstract}
We report on recent extensions and improvements to the {\legolas} code, which is an open-source, finite element-based numerical framework to solve the linearised (magneto)hydrodynamic equations for a three-dimensional force- and thermally balanced state with a nontrivial one-dimensional variation. The standard Fourier modes imposed give rise to a complex, generalised non-Hermitian eigenvalue problem which is solved to quantify all linear wave modes of the given system in either Cartesian or cylindrical geometries. The framework now supports subsystems of the eight linearised MHD equations, allowing for pure hydrodynamic setups, only one-dimensional density/temperature/velocity variations, or the option to treat specific closure relations. We discuss optimisations to the internal datastructure and eigenvalue solvers, showing a considerable performance increase in both execution time and memory usage. Additionally the code now has the capability to fully visualise eigenfunctions associated with given wave modes in multiple dimensions, which we apply to standard Kelvin-Helmholtz and Rayleigh-Taylor instabilities in hydrodynamics, thereby providing convincing links between linear stability analysis and the onset of non-linear phenomena.
\end{abstract}


\end{frontmatter}

\section{Introduction} \label{sect: introduction} \noindent
The {\legolas} code\footnote{Open-source under the GPL-3.0 license and available on GitHub at \url{https://github.com/n-claes/legolas}} is an open-source numerical framework capable of solving the full set of linearised (magneto)hydrodynamic equations for a three-dimensional force- and thermally balanced equilibrium in either Cartesian or cylindrical geometries, assumed to have a non-trivial variation along one dimension \cite{claes2020_legolas,dejonghe2022}. Standard Fourier modes are imposed on the other coordinates. {\legolas} supports the inclusion of background flows, external gravity, anisotropic thermal conduction, resistivity, optically thin radiative losses, viscosity, and full Hall MHD; allowing for a treatment of highly generalised setups.
The applied Fourier analysis gives rise to a set of partial differential equations, which are reduced further by relying on a finite element representation combined with the weak Galerkin formalism to approximate the unknown functions by a linear combination of cubic and quadratic basis functions on a spatially discretised grid. This results in a generalised, non-Hermitian complex eigenvalue problem which can be solved to yield all linear modes of the given equilibrium, including their eigenfunctions, allowing for in-depth studies of system stability and wave mode behaviour.

The general study of plasma instabilities is of fundamental importance to a wide range of applications, ranging from laboratory conditions to general astrophysical systems. Over the past few decades various numerical tools have been constructed for exactly that purpose, with early examples being \textsf{LEDA} \cite{kerner1985} and \textsf{LEDAFLOW} \cite{nijboer1997}, in which {\legolas} has its early roots. Extending the basic principles behind these kind of codes to higher dimensions is highly relevant in the scope of fusion research, where it is of paramount importance to properly classify possible waves and instabilities of general Grad-Shafranov equilibria. Examples of higher-dimensional codes include \textsf{CASTOR} \cite{kerner1998}, which calculates normal modes in resistive MHD for 2D tokamak configurations; its extension \textsf{STARWALL} \cite{merkel2015}, capable of studying linear stability of 3D ideal equilibria with resistive conducting structures; and the \textsf{SPEC} code \cite{hudson2012,kumar2021}, which computes linear instabilities of Multi-Region Relaxed MHD (MRxMHD) in 3D toroidal and cylindrical geometries. A slightly different approach is used by \textsf{MINERVA} \cite{aiba2009}, which solves the Frieman-Rotenberg equation as both an initial value problem and a generalised eigenvalue problem to investigate the effect of toroidal rotation on the stability of ideal MHD modes in tokamak configurations. More recently \textsf{SCELT} was developed in \cite{ma2022}, which relies on symbolic computation and automatic numerical discretisation in order to construct the matrix coefficients in the eigenvalue problem, and solves for linear stability of resistive plasmas in toroidal geometry.

Numerical astrophysics is of course not limited to these linear codes, and over the years a myriad of various nonlinear (M)HD frameworks have been developed, capable of performing large-scale multi-dimensional simulations of numerous (astro)physical phenomena. These codes usually rely on finite volume methods and block-based grid structures, allowing for massively parallelised solver capabilities across a large number of cores. Examples include \textsf{MPI-AMRVAC}, a modular, parallel adaptive mesh refinement (AMR) framework for solving hyperbolic partial differential equations \cite{porth2014_amrvac,xia2018_amrvac,keppens2023_amrvac}; \textsf{FLASH}, a high-performance AMR multi-physics application code \cite{fryxell2000_flash}; \textsf{PLUTO}, targeting high-mach number flows in astrophysical fluid dynamics on AMR grids; \textsf{Athena++}, a general relativistic MHD code and AMR framework; \textsf{ENZO}, a community-driven AMR code tailored to multi-physics hydrodynamic astrophysical applications \cite{bryan2014_enzo}; and the Black Hole Accretion Code \textsf{BHAC}, capable of solving the ideal general relativistic MHD equations in multiple dimensions on arbitrary stationary spacetimes using an efficient parallel block-based approach \cite{porth2017_bhac,olivares2019_bhac}.

The aforementioned linear codes are quite generalised for each of their specific application areas, though they commonly contain only a limited number of physical effects that are included, usually treating resistive/viscous/conductive MHD or combinations thereof, mostly in tokamak-like geometries. While {\legolas} only treats 3D plasmas with 1D variation (for now), the various supported physical effects ensure a wide range of applicability across various configurations. Since its creation it has been applied to viscous and Hall MHD \cite{dejonghe2022}, to investigate thermal instabilities in non-adiabatic solar atmosphere settings \cite{claes2021}, to verify eigenvalue solutions in magnetic flux tubes with background rotational flow \cite{skirvin2023}, and to verify results on the Super Alfv\'enic Rotational Instability (SARI) in accretion disks \cite{goedbloed2022}. Due to {\legolas}'s high modularity it can straightforwardly be extended to include additional physics and geometries in the future. Furthermore, with the extensions and improvements reported upon in this work the code can support different sets of equations as well, further widening its already broad application range.

The previous version of {\legolas} as introduced in \cite{claes2020_legolas} focused more on the physical applications themselves, comparing results against known literature. Consequently, not much attention was given to performance optimisations since code validation and testing took priority. {\legolasOne} was solving the eigenvalue problem by using a QR-algorithm based on LAPACK routines \cite{book_lapack}, resulting in all eigenvalues with corresponding eigenfunctions. Solvers based on an implicitly restarted Arnoldi iteration through ARPACK \cite{book_arpack} to calculate a specific number of eigenvalues with particular features were also available. This was all done in a dense manner, where the entire matrices were stored in memory. While this was convenient to work with implementation-wise, it is inefficient, slow, and unfeasable for large-scale applications that employ matrix dimensions of tens of thousands of elements, especially due to the sparse nature of the eigenvalue problem. To study for example continuum modes, which are inherently singular solutions to the linear partial differential equations, extreme resolutions are needed to attempt a proper approximation of their singular and ultra-localised eigenfunctions.
Here we report on {\legolasTwo}, which is characterised by major improvements and extensions to the existing solvers along with a complete overhaul of the internal datastructure, yielding considerable performance gains.

Section \ref{sect: structure} contains a brief discussion of the generalised eigenvalue problem and how the matrices are internally handled. This is followed by a detailed overview of all available solvers in Section \ref{sect: solvers}, including their approaches in reducing the eigenvalue problem to standard form and the various steps taken to reach the solutions. In Section \ref{sect: performance} the various improvements and extensions are benchmarked against previous {\legolasOne} results, discussing performance gains as a function of resolution for each of the available solvers. Section \ref{sect: subsystems} showcases a new major feature of the framework, which is the capability to solve subsets of the general set of eight MHD equations, allowing for pure hydrodynamics or specific density/velocity/temperature variations in 1D or 2D. This functionality is applied to specific hydrodynamic setups susceptible to the well-known Kelvin-Helmholtz and Rayleigh-Taylor instabilities, discussing their spectra and the eigenfunctions of unstable modes. These eigenfunctions are then used in Section \ref{sect: visualisations} to highlight an additional new feature of {\legolasTwo}, one that allows for eigenmode visualisations in multiple dimensions. The resulting figures provide convincing links between linear stability analysis and the onset of expected non-linear phenomena, opening the door to in-depth studies of the possible link between non-linear features and linear theory.

\section{Background \texorpdfstring{\&}{and} Internal datastructure} \label{sect: structure} \noindent

\subsection{Generalised eigenvalue problems} \label{sect: generalised_evp} \noindent
In the general case {\legolas} treats the full set of eight (magneto)hydrodynamic equations, which are linearised around a 3D background with a non-trivial one-dimensional variation. Standard Fourier modes are imposed on the other coordinates $u_2$ and $u_3$ which resolve to $x$ and $y$ in Cartesian geometries and $\theta$ and $z$ in cylindrical geometries. These standard modes have an exponential time dependence and are of the form
\begin{equation} \label{eq: fourier}
  f_1 = \hat{f}_1(u_1)\exp\left[i\left(k_2 u_2 + k_3 u_3 - \omega t\right)\right],
\end{equation}
where $\omega = \wreal + i\wimag$ represents the complex frequency of the wave modes and $f_1$ denotes any of the perturbed quantities. As such, instabilities are characterised by a positive $\wimag$, while stable (damped) modes have $\wimag < 0$. In Cartesian geometries the spatial coordinate $u_1$ and wave numbers $k_2$ and $k_3$ are given
by $x$, $k_y$ and $k_z$, respectively; for cylindrical geometries these are $r$, $m$, and $k$, with $m$ an integer quantity due to periodicity.

The set of linearised equations itself will not be given here, instead we refer to the appendices in \cite{dejonghe2022}. These are still differential equations, which are transformed further by relying on a finite element representation, resulting in a generalised, non-Hermitian complex eigenvalue problem of the form
\begin{equation} \label{eq: evp}
  \amat \bfx = \omega \bmat \bfx,
\end{equation}
where $\bfx$ denotes the state vector of perturbed quantities, given by
\begin{equation} \label{eq: state_vector}
  \bfx = \left(\rho_1, v_1, v_2, v_3, T_1, a_1, a_2, a_3\right)^\top.
\end{equation}
Here $\rho_1$ denotes the density, $\bfv_1$ the velocity vector, $T_1$ the temperature and $\bfa_1$ the magnetic vector potential, from which the perturbed magnetic field can be obtained using the curl operator. The matrix $\amat$ is generally speaking complex and non-Hermitian, the $\bmat$-matrix is real and in most cases symmetric. Some physical effects break $\bmat$-matrix symmetry if they are included, for example the electron inertia quantities in the Hall terms \cite{dejonghe2022}.

\subsection{Matrix storage} \label{sect: matrix_storage} \noindent
As discussed in \cite{claes2020_legolas} both matrices of the eigenvalue problem are block-tridiagonal, where the size of a block is connected to the number of linearised equations that are being solved. Generally speaking, in full MHD each block will have a size of $16 \times 16$ (eight equations with a factor two due to the finite element treatment) and at most three blocks appear next to each other, resulting in a maximum of 31 sub- or superdiagonals. Furthermore the blocks themselves can be either dense or sparse, depending on the physical effects taken into consideration.

\begin{figure*}[t]
  \centering
    \includegraphics[width=\textwidth]{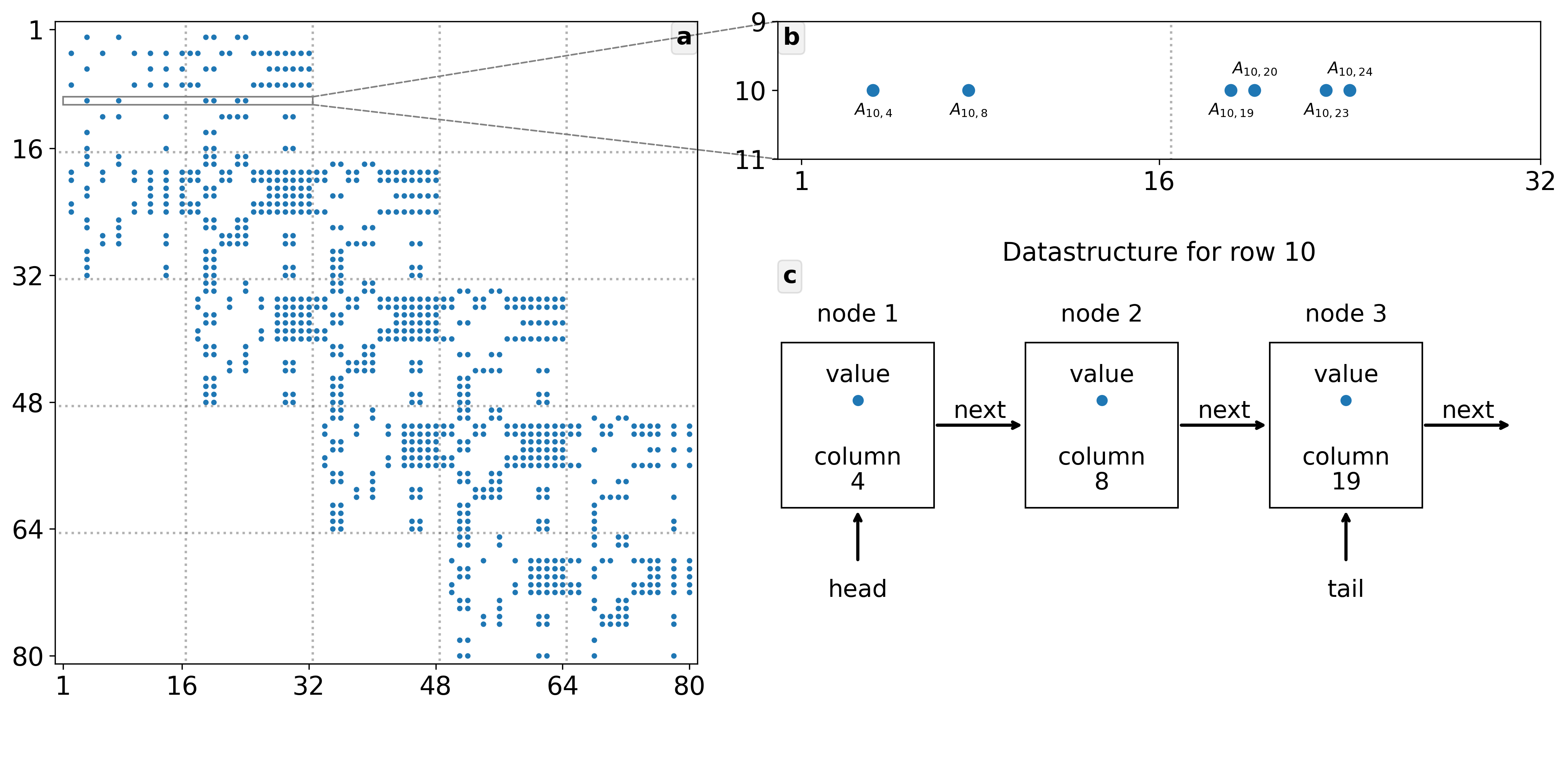}
    \caption{
      Panel a: a general {\legolas} matrix for five base gridpoints in MHD. Every dot corresponds to a non-zero value.
      Panel b: zoom-in of the tenth row of the matrix, showing the six non-zero elements.
      Panel c: custom linked-list datastructure for (part of) the specific matrix row.
    }
  \label{fig: structure}
\end{figure*}

Instead of relying on a standard Compressed Sparse Row (CSR/Yale) or Compressed Sparse Column (CSC) format we make use
of a custom built linked list implementation, the reason for this being the need for a highly flexible datastructure that can easily be modified during or after the assembly process. The sparsity of the blocks themselves is problem-dependent: for large-scale applications in ideal (magneto)hydrodynamics, for example, the blocks are very sparse such that over half of the elements inside the band are zero, and generally speaking the blocks get denser when more physical effects are taken into account. The implementation of the boundary conditions requires removal or addition of elements (and even whole rows and/or columns) after the matrices have already been assembled. How many non-zero elements are being modified is difficult to know a priori before matrix assembly, since it depends on the users' choice of physics, boundary conditions, and selected basis functions. This kind of dynamic assembly is not straightforward to do for fixed band storage (i.e. CSC/CSR) and requires detailed tracking of row and column indices that are being modified and possibly one or more reallocations of the matrix arrays in memory. The chosen linked list approach is flexible, easy to use, and circumvents the need for memory reallocations. Additionally, after the matrices have been fully assembled it would certainly be possible to convert them into a more conventional CSR/CSC format for further processing if demanded by future applications or extensions.

Figure \ref{fig: structure} shows a visualisation of how a matrix is stored in memory using the linked list representation. Panel \textbf{a} depicts the typical structure of a {\legolas} matrix, shown here using only five base gridpoints for visual clarity. The matrix itself is $80 \times 80$ and has 886 non-zero elements in total corresponding to the blue dots. Panel \textbf{b} gives a closer look at the tenth row in the matrix, showing the (in this case) six non-zero elements. The custom linked-list datastructure is illustrated in Panel \textbf{c}. The entire matrix is reduced to a single array where each index corresponds to a row of the matrix, represented as a linked list consisting of nodes. Every node contains three items: a non-zero matrix element, the corresponding column index, and a pointer pointing towards either the next node in the list, or \textsf{null} in case there are no further nodes. Every linked list itself keeps track of the total number of nodes in the list and additionally has two pointers, \textsf{head} and \textsf{tail}, pointing towards the first and last element in the list. Both of these point to the same node if there is only one present, or both to \textsf{null} if the list is empty. Traversing through the list is straightforward, starting from \textsf{head} and following \textsf{next} until the last node is reached.

Retrieving an element $A_{ij}$ happens by accessing the main array at index $i$, then traversing the list until the column index of a node matches the required index $j$, followed by retrieval of the associated element. Inserting or  removing nodes from the list is also straightforward, that is, adding an additional node somewhere in the list and updating the corresponding pointers. This makes handling of for example the essential boundary conditions very convenient, where usually one or more rows and columns are removed from the matrix as discussed in
\cite{claes2020_legolas}. The datastructure as implemented here also makes matrix-vector multiplication a very cheap operation, which is nothing more than traversing through each list, multiplying every node with the corresponding vector element and summing them. Since no zero elements are stored in the list only the necessary amount of operations is performed to obtain the result. This is particularly useful for iterative solvers such as ARPACK \cite{book_arpack}, which in most cases purely rely on matrix-vector products.
\clearpage

\section{Eigenproblem solvers} \label{sect: solvers} \noindent
\subsection{QR-based solvers} \noindent
Initially {\legolas} was handling the eigenvalue problem \eqref{eq: evp} by reducing it to standard form through
$\bmat^{-1}\amat\bfx = \omega \bfx$. In obtaining the left-hand side operator the $\bmat$-matrix was explicitly inverted, the product was calculated and the result passed on to LAPACK. As the $\bmat$-matrix is never singular this is always an option, however, the inversion of a matrix is an expensive operation and furthermore this implied storing three dense matrices in memory ($\amat$, $\bmat$ and the inverse product). Clearly this does not scale at all to high resolutions with matrix sizes of say a few thousand and higher, without even considering numerical inaccuracies arising from inversion and multiplication.

Here we introduce updates to these solvers, which are faster and require (much) less memory.
The two matrices $\amat$ and $\bmat$ are now stored using the datastructure described in Section \ref{sect: structure}. Unfortunately, LAPACK still requires passing a dense matrix to its solution routine for generalised, non-Hermitian eigenvalue problems, and the eigenvalue problem still needs reduction to standard form. This is now done using two approaches.

\paragraph{a) LU factorisation} Here the reduction to standard form still relies on $\bmat^{-1}\amat$, which is now obtained using the following steps:
\begin{enumerate}
  \item The $\bmat$-matrix is converted from its linked-list datastructure to banded storage, while $\amat$ is converted to a dense representation.
  \item The system of linear equations $\bmat\bfx = \amat$ is solved for $\bfx$ using a direct linear system solver,
  thereby relying on a standard LU decomposition with row interchanges and partial pivoting to factor
  $\bmat$ as $\bmat = \Lmat\Umat$. Here $\Lmat$ and $\Umat$ are lower and upper triangular matrices, respectively.
  \item The previous point yields $\bmat^{-1}\amat$ without explicitly inverting the $\bmat$-matrix, this product is then passed on to LAPACK to solve for all eigenvalues and eigenvectors.
\end{enumerate}

\paragraph{b) Cholesky factorisation} In most cases the $\bmat$-matrix is symmetric and thus Hermitian positive definite. This means that a Cholesky factorisation of $\bmat$ can be calculated, which is given by $\bmat = \Umat\Umat^\hermconj$.
Here $\Umat$ is an upper triangular matrix and $\Umat^\hermconj$ denotes the complex conjugate of $\Umat$. The reduction to standard form can now be written as $\Umat^{-\hermconj}\amat\Umat^{-1}\bfx = \omega\bfx$, obtained through the following steps:
\begin{enumerate}
  \item The $\bmat$-matrix is converted from its linked-list datastructure to upper Hermitian banded storage, while $\amat$ is converted to a dense representation.
  \item The Cholesky factorisation $\bmat = \Umat^\hermconj\Umat$ is calculated.
  \item The system $\Umat^\hermconj\bfx = \amat$ is solved for $\bfx$, yielding $\Umat^{-\hermconj}\amat$.
  \item The system $\bfx\Umat = \Umat^{-\hermconj}\amat$ is solved for $\bfx$, yielding
  $\Umat^{-\hermconj}\amat\Umat^{-1}$.
  \item The result is passed on to LAPACK to solve for all eigenvalues and eigenvectors.
\end{enumerate}
Using the Cholesky decomposition is both more numerically stable and more efficient, however, this solution method can not be used if physical terms are taken into account that break $\bmat$-matrix symmetry.

Both approaches described here only require us to store one dense matrix representation in memory instead of three. Furthermore, since most calculations use LAPACK's banded matrix routines this is (much) more efficient than their dense counterparts, significantly speeding up computational time. It should be noted that these QR-based solvers still do not scale well to very large matrices as the final call to LAPACK's solver still requires a dense matrix. Banded variants of these QR solvers will not perform any better, as they typically require a large amount of iterations and quickly loose the banded structure. For example, a matrix of size $N \times N$ with a band of size $K$ will have, after the Hessenberg upper triangular matrix reduction, a band size of approximately $K^2$. Afterwards every QR iteration will increase the band size by one. Usually these algorithms need around $3N$ iterations, meaning the sparse banded structure will quickly be lost and the approach becomes unfeasible for very large matrices.

\subsection{Implicitly restarted Arnoldi methods} \label{sect: arnoldi} \noindent
Whereas the QR based algorithms are direct methods able to calculate the full spectrum, Arnoldi-based algorithms are iterative methods to calculate a select number of eigenvalues with specific properties (largest magnitude, smallest imaginary part, etc.). The main advantage here is that these methods only need matrix-vector products during their iterative process, such that all sparsity properties can be exploited to the fullest which makes them perfectly suited for large-scale sparse eigenvalue problems. {\legolasOne} also had these methods implemented, but due to the use of dense matrices they were very ineffective and slow. The new datastructure as discussed in Section \ref{sect: structure} has very efficient matrix-vector products, such that the new implementations in {\legolasTwo} lead to a huge performance boost, with a significant decrease of memory usage (no dense matrices need to be stored) and computational time. Two Arnoldi variants are implemented.

\paragraph{a) General Arnoldi solver}
The first variant is the Arnoldi-variant of the QR-methods described earlier, that is, solving the standard eigenvalue problem $\bmat^{-1}\amat\bfx = \omega\bfx$. However, during the iteration only the matrix-vector product on the left-hand side of this equation is required, i.e. $\Rvec = \bmat^{-1}\amat\bfx$. In order to obtain this the following steps are taken:
\begin{enumerate}
  \item The $\bmat$-matrix is converted from its linked-list structure to a banded matrix.
  \item Iteration starts.
  \item Calculate the vector $\uvec = \amat\bfx$.
  \item Solve the linear system $\bmat \Rvec = \uvec$ for the vector $\Rvec$, using the same direct linear system solver as for the QR methods (that is, using the LU decomposition). Stop iteration when eigenvalues are converged.
\end{enumerate}

\paragraph{b) Shift-invert spectral transformation}
The second variant relies on a spectral transformation, which allows one to select a complex shift $\sigma$ around which a given number of eigenvalues are calculated. This particular method enhances convergence for a specific part of the spectrum and relies on the transformation $\omega \rightarrow \sigma + 1/\lambda$. This transforms the original eigenvalue problem \eqref{eq: evp} into
\begin{equation} \label{eq: transformed_evp}
  \Gmat\bfx = \lambda\bfx  \quad\text{with}\quad \Gmat \equiv \left(\amat - \sigma\bmat\right)^{-1}\bmat.
\end{equation}
It follows from the applied transformation
\begin{equation}
  \lambda = \frac{1}{\omega - \sigma},
\end{equation}
that finding the eigenvalues $\lambda_\itxt$ with largest magnitude corresponds to the eigenvalues $\omega_\itxt$ of the original problem that are closest to the chosen shift $\sigma$ in absolute value. Equation \eqref{eq: transformed_evp} is again a standard eigenvalue problem, such that again the matrix-vector product $\Rvec = \Gmat\bfx$ is needed:
\begin{enumerate}
  \item The LU-decomposition of $\amat - \sigma\bmat$ is calculated and stored in banded form.
  \item Iteration starts.
  \item Calculate the vector $\uvec = \bmat\bfx$.
  \item Solve the linear system $\left(\amat - \sigma\bmat\right)\Rvec = \uvec$ for the vector
  $\Rvec$, reusing the LU factorisation of $\amat - \sigma\bmat$ calculated earlier. Stop iterating when eigenvalues
  are converged.
  \item Eigenvalues are retransformed to those of the original problem using
  $\omega_\itxt = \sigma + 1/\lambda_\itxt$.
\end{enumerate}
While these approaches still require us to store one matrix in memory as it is needed for the linear system solver, this can now be done using solely a banded matrix instead of a dense one.

\subsection{Inverse iteration} \label{sect: inverse-iteration}\noindent
The process of inverse iteration takes a user-defined shift $\sigma$ and iterates to find an approximate eigenvector. {\legolasTwo} implements an extension to this method, known as the Rayleigh quotient procedure, to obtain an increasingly accurate eigenvalue after each iteration. Here we will give a brief overview of the various steps taken in the implementation, but will not go into detail on the method itself. For an extensive discussion on iterative methods, including the Rayleigh quotient iteration, we refer to \cite{book_demmel} or \cite{book_trefethen}.
The iteration proceeds as follows:
\begin{enumerate}
  \item The LU-decomposition of $\amat - \sigma\bmat$ is calculated and stored in banded form, the eigenvalue
        $\mu_\itxt$ is set to the provided shift $\sigma$.
  \item The system $\Umat\bfx = \boldsymbol{1}$ is solved for $\bfx$ to provide an initial eigenvector guess
        $\bfx_\itxt$.
  \item Iteration starts.
  \item The Rayleigh quotient is used to provide a better eigenvalue approximation through
        \begin{equation}
          \mu_\ipone = \frac{\bfx_\itxt^\hermconj \amat \bfx_\itxt}{\bfx_\itxt^\hermconj \bmat \bfx_\itxt}.
        \end{equation}
  \item The eigenvalue is converged if
        \begin{equation} \label{eq: inverse_criterion}
          \lVert \amat\bfx_\itxt - \mu_\ipone\bmat \bfx_\itxt\rVert < \lvert \mu_\ipone \rvert \epsilon,
        \end{equation}
        with $\epsilon$ a user-defined tolerance. Typical values of $\epsilon \approx 10^{-10}$ are more than sufficient. Stop iterating if this condition is satisfied, otherwise continue to the next step.
  \item A better eigenvector approximation is obtained through
        \begin{equation}
          \bfx_\ipone = \frac{\left(\amat - \sigma\bmat\right)^{-1}\bmat\bfx_\itxt}{
            \lVert \left(\amat - \sigma\bmat\right)^{-1}\bmat\bfx_\itxt \rVert
          },
        \end{equation}
        where the numerator is calculated by solving the linear system
        $\left(\amat - \sigma\bmat\right)\bfx_\itxt = \bmat\bfx_\itxt$ for $\bfx_\itxt$ using the LU-decomposition calculated earlier. Continue the iteration, use $\mu_\ipone$ and $\bfx_\ipone$ in the next step.
\end{enumerate}
Inverse iteration usually converges much faster than the earlier discussed shift-invert. Generally speaking, the closer the shift $\sigma$ is to the actual eigenvalue $\omega$, the faster the procedure will converge. Note that only one eigenvalue is calculated during the iteration and that only matrix-vector products are needed, which makes this method perfectly suitable to analyse isolated eigenvalues at high resolutions. If the approximate locations of the desired eigenvalues are known a priori (e.g. through a low-resolution run using QR-variants) multiple iterations with different shifts can be performed to compose a local spectrum.

\section{Benchmarks and performance} \label{sect: performance} \noindent
In this Section we benchmark the various improvements and solvers discussed earlier against the previous
{\legolasOne} version, looking at both computational time and memory consumption. All runs are performed on a desktop running Rocky Linux 8.7 with 128 GB of memory and Intel{\reg} Xeon{\reg} Silver 4210 (2.20GHz) CPU's. Compilation of {\legolas} was done using \textsf{GCC 12.1.0} with \textsf{-03} optimisation flags enabled, \textsf{OpenBLAS 0.3.15} was used for the libraries. Benchmarking itself was done using Python, by spawning a {\legolas} subprocess with a memory profiler attached.

The background used in all cases is taken from \cite{goedbloed2018} and considers an accretion disk setup containing non-trivial density and temperature profiles in a magnetised disk with background rotational flow and radially directed gravitational profile, susceptible to magnetorotational instabilities. The reason for this particular setup is threefold. First, \cite{goedbloed2018} calculated the spectrum using the spectral web approach, a different method than the eigenvalue problem approach {\legolas} uses, which makes it an ideal testcase to compare eigenvalues. Second, the magnetised accretion disk contains background flow and has a gravitational profile, which makes sure that the blocks in our matrices are not too sparse. Finally, the setup contains a relatively large sequence of isolated instabilities that are properly resolved at low resolutions (50 gridpoints already resolves most of the unstable sequence). This ensures that a proper shift can be chosen in methods like shift-invert and inverse iteration such that they can converge towards one of the isolated eigenvalues in the sequence.

\subsection{Reference equilibrium} \label{sect: benchmark_bg} \noindent
\begin{figure*}[t]
  \centering
  \includegraphics[width=\textwidth]{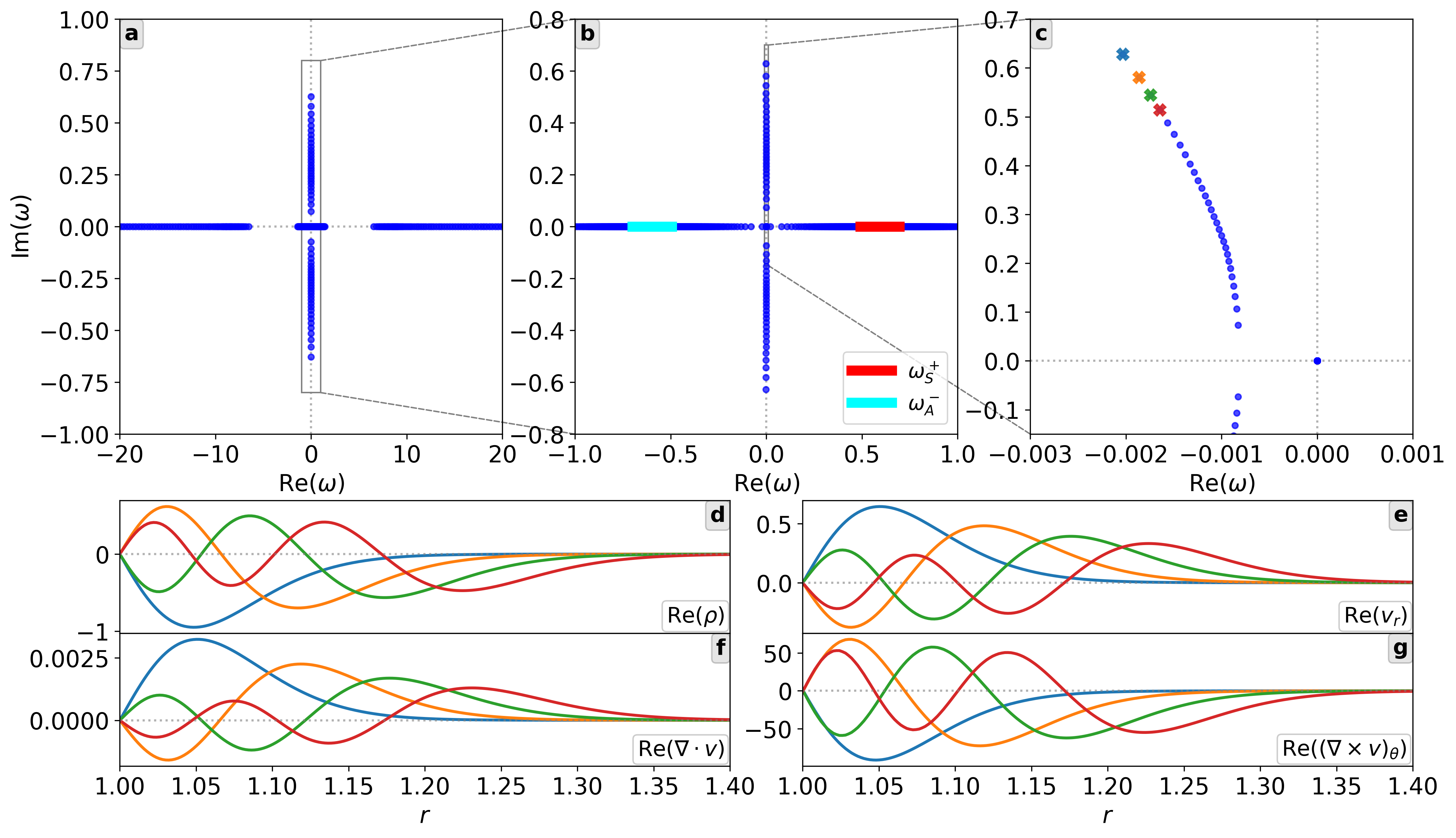}
  \caption{
    Spectrum and eigenfunctions for the reference equilibrium \eqref{eq: mri_bg}, obtained with QR-Cholesky at 250 gridpoints. Panel a: (part of) the spectrum at larger scales, panels b and c zoom in further.
    The slow $(\omega_S^+)$ and Alfv\'en $(\omega_A^-)$ continua are indicated on panel b.
    Panels d through g: real eigenfunctions for the perturbed density, radial velocity component, velocity divergence and $\theta$-component of the velocity curl, respectively; for the four eigenmodes annotated on panel c.
  }
  \label{fig: full_mri_spectrum}
\end{figure*}
For convenience we give the equilibrium background here, for an extensive discussion we refer to \cite{goedbloed2018}. The geometry is cylindrical in $r \in [1, 2]$, with background profiles given by
\begin{equation} \label{eq: mri_bg}
  \begin{gathered}
    \rho_0(r) = r^{-3/2}, \quad T_0(r) = \varepsilon^2 r^{-1}, \\
    v_{0\theta}(r) = \Omega_1 r^{-1/2}, \\
    B_{0\theta}(r) = \mu B_c r^{-5/4}, \quad B_{0z}(r) = B_c r^{-5/4}.
  \end{gathered}
\end{equation}
It is straightforward to check that this system is force balanced, provided we ensure the following:
\begin{equation}
  \Omega_1^2 = 1 - \frac{5}{2}\varepsilon^2 - \frac{1}{2}\left(\mu^2 - \frac{5}{2}\right)B_c^2.
\end{equation}
Here, $\Omega_1$ is the angular rotation parameter, $\varepsilon$ relates to pressure magnitude, and the magnetic field constant is denoted by
\begin{equation}
  B_c^2 = \frac{2\varepsilon^2}{\beta(\mu^2 + 1)}.
\end{equation}
The entire system is hence completely determined by choosing the parameters $\mu$, $\varepsilon$ and $\beta$. In all cases that follow these have been set to unity, $0.1$, and $100$, respectively; the wavenumbers are equal to $k_2 = m = 0$ and $k_3 = k_z = 70$. All quantities are in normalised (dimensionless) units.

Figure \ref{fig: full_mri_spectrum} shows the spectrum in the complex $\omega$-plane for the reference equilibrium given by Equation \eqref{eq: mri_bg}. Panel \textbf{a} shows (part of) the spectrum at larger scales, Panel \textbf{b} zooms in further near the origin, with the forwards slow and backwards Alfv\'en continua $\omega_S^+$ and $\omega_A^-$ annotated with red and cyan bands, respectively; both lying on the real axis. Panel \textbf{c} zooms in on the unstable sequence, where it is important to stress that for this particular setup there are only a finite amount of instabilities present: here we have 29 actual unstable eigenvalues with the rest of the (infinite) sequence being stable, matching the Spectral Web results given in \cite{goedbloed2018}. Panels \textbf{d} through \textbf{g} of Figure \ref{fig: full_mri_spectrum} show the real part of the eigenfunctions corresponding to the four eigenvalues annotated with a cross in Panel \textbf{c}. The perturbed part of the density is shown (panel \textbf{d}), along with the radial velocity component (panel \textbf{e}), velocity divergence (panel \textbf{f}) and $\theta$-component of the velocity curl (panel \textbf{g}) as a function of radius. Note that for these four panels only a part of the domain is shown, as all eigenfunctions are rather localised and approach zero for larger radii.
Table \ref{tab: mri_eigenvalues} contains the actual eigenvalues for the first 10 instabilities in the unstable branch, up to eight digits of precision, for reproduction purposes. The spectrum and eigenfunctions were calculated using QR-Cholesky at 250 gridpoints.

\begin{table}[t!]
  \centering
  \caption{
    Eigenvalues of the first 10 instabilities in Figure \ref{fig: full_mri_spectrum}, panel c, corresponding to the equilibrium \eqref{eq: mri_bg}. Values obtained using QR-Cholesky at 250 gridpoints.
  }
  \begin{tabular}{cc}
    \T
    Mode & Eigenvalue \\
    \hline
    \T
    $\omega_1$ & $-0.00203122 + 0.62772161 \itxt$ \\
    $\omega_2$ & $-0.00186300 + 0.58048727 \itxt$ \\
    $\omega_3$ & $-0.00174203 + 0.54438082\itxt$ \\
    $\omega_4$ & $-0.00164567 + 0.51413241 \itxt$ \\
    $\omega_5$ & $-0.00156509 + 0.48768537 \itxt$ \\
    $\omega_6$ & $-0.00149570 + 0.46396379 \itxt$ \\
    $\omega_7$ & $-0.00143472 + 0.44231495 \itxt$ \\
    $\omega_8$ & $-0.00138033 + 0.42230484 \itxt$ \\
    $\omega_9$ & $-0.00133127 + 0.40362607 \itxt$ \\
    $\omega_{10}$ & $-0.00128659 + 0.38605047 \itxt$ \\
  \end{tabular}
  \label{tab: mri_eigenvalues}
\end{table}

\subsection{Performance of QR} 
\begin{figure}[t]
  \centering
  \includegraphics[width=\columnwidth]{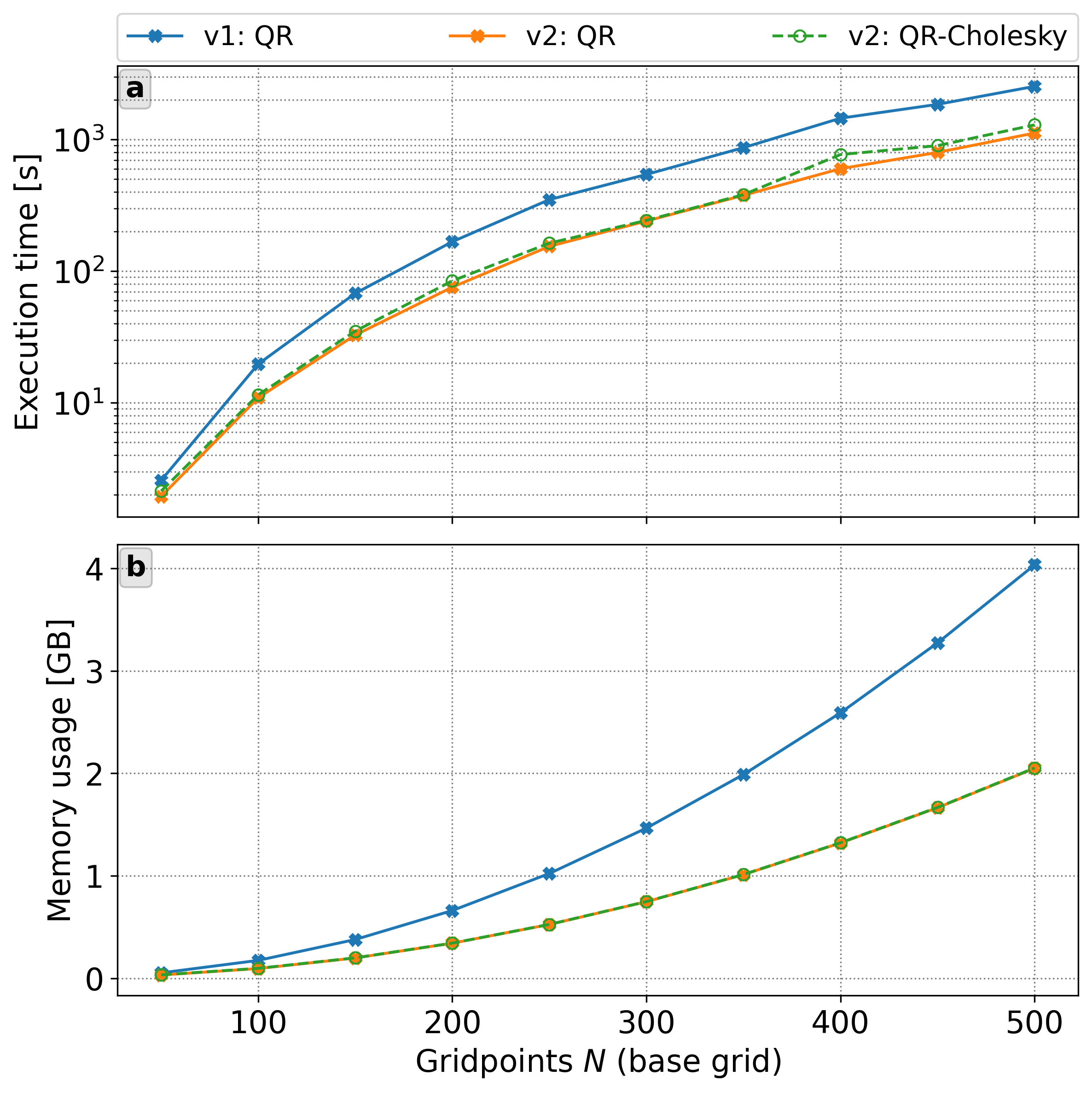}
  \caption{
    Execution time (Panel a) and memory usage (Panel b) for the QR-based solvers. The blue curve represents
    {\legolasOne}, the orange and green ones {\legolasTwo}. In all cases there is approximately a 50\%
    increase in performance.
  }
  \label{fig: qr_benchmarks}
\end{figure}

\noindent
First we benchmark the performance of both QR-based algorithms using an increasing number of gridpoints $N$ for the base grid. The background discussed in the previous subsection is full MHD, hence the matrix dimensions correspond to $16N \times 16N$. Figure \ref{fig: qr_benchmarks} shows the execution time in seconds (panel \textbf{a}) and the maximum memory usage in GB (panel \textbf{b}) as a function of $N$, for every $N$ all three versions yield $16N$ complex eigenvalues with corresponding eigenvectors. The blue curve represents QR with explicit inversion in {\legolasOne}, the solid orange and green dashed curves represent QR-LU and QR-Cholesky in {\legolasTwo}, respectively, with both curves approximately overlapping. For 500 gridpoints the runtime decreases from $\approx 2500$ seconds to $\approx 1100$ seconds, a performance increase of $\approx 56\%$. QR-Cholesky is slightly slower at higher resolutions (roughly $+200$ seconds at 500 gridpoints), most likely due to the fact that two linear systems need to be solved compared to only one for QR-LU, though it should be more stable numerically. The maximum memory usage approximately halves in all cases. The eigenvalues obtained for all three cases are identical up to absolute differences of approximately $10^{-12}$ or less.

\subsection{Performance of shift-invert} \noindent
\begin{figure}[t]
  \centering
  \includegraphics[width=\columnwidth]{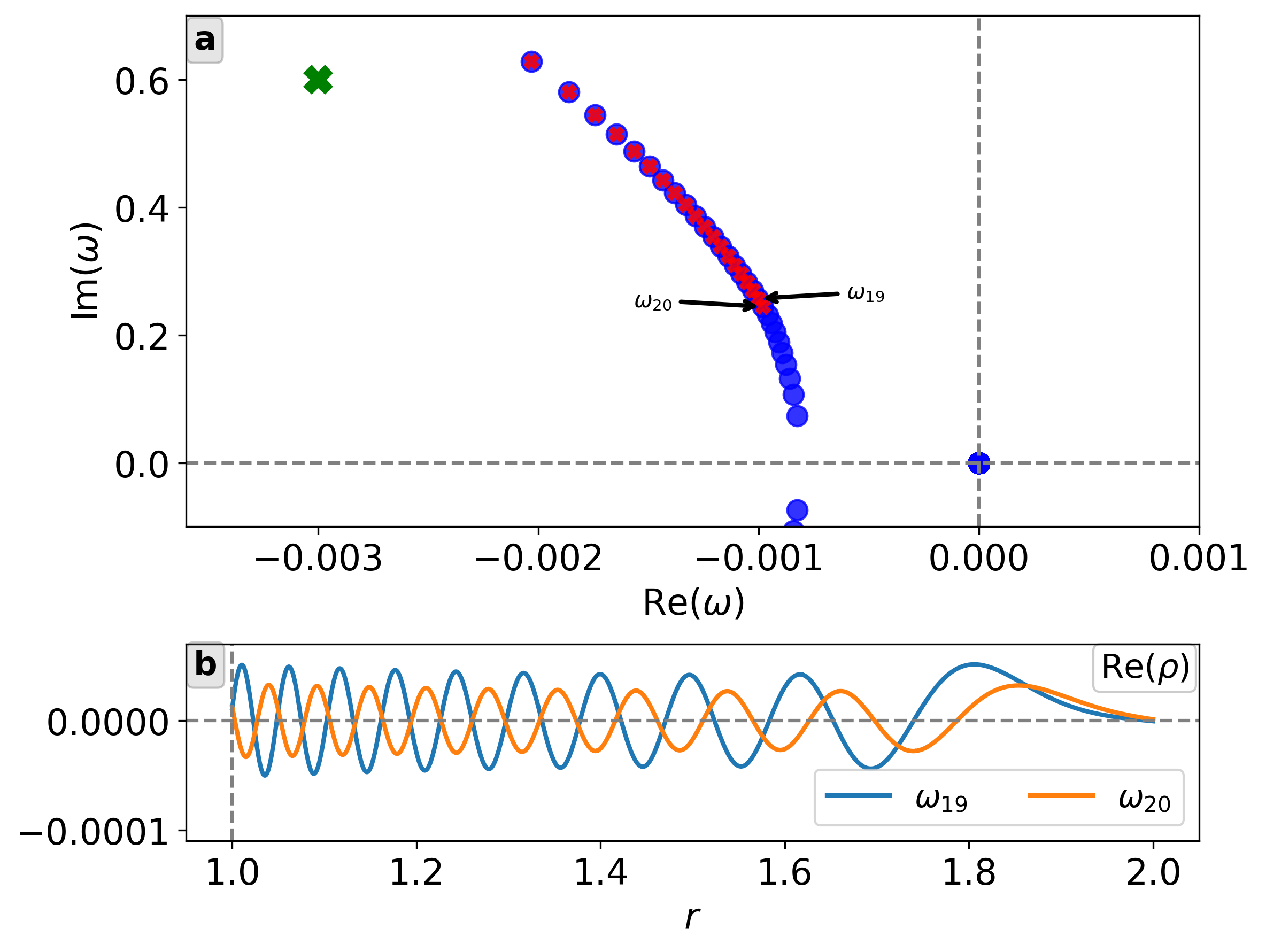}
  \caption{
    Panel a: partial spectrum for the accretion disk in \eqref{eq: mri_bg}, blue dots denote eigenvalues obtained by a reference run with QR-Cholesky at 250 gridpoints. Red crosses denote the 20 eigenvalues obtained by shift-invert at 2000 gridpoints with $\sigma = -0.003 + 0.6\itxt$ (green cross). Panel b: real part of the $\rho$ eigenfunctions of the 19th and 20th mode in the sequence, annotated with arrows on Panel a.
  }
  \label{fig: spectrum_si}
\end{figure}
While the performance increase of the QR-solvers is rather significant in version two compared to version one, the fact that one dense matrix still has to be stored in memory is clearly still a bottleneck. For the iterative methods on the other hand this requirement is lifted and a much higher performance difference is expected. Here we benchmark the same setup as discussed in subsection \ref{sect: benchmark_bg}, now using ARPACK's shift-invert method as explained in subsection \ref{sect: arnoldi}. For the shift we take $\sigma = -0.003 + 0.6\itxt$ and query for $20$ eigenvalues with largest magnitude. Figure \ref{fig: spectrum_si} shows (a part of) the spectrum associated with the chosen setup, that is, the aforementioned magnetised accretion disk in \eqref{eq: mri_bg}. The reference spectrum obtained with QR-Cholesky at 250 gridpoints is shown using blue dots, the $20$ eigenvalues obtained through shift-invert at 2000 gridpoints are annotated with red crosses. The shift itself is denoted with a green cross and results in a calculation of the first $20$ eigenvalues in the unstable branch, matching the results in Table \ref{tab: mri_eigenvalues} up to absolute differences of roughly $10^{-10}$. A visual comparison can be made with Figure 13 from \cite{goedbloed2018}, indicating excellent correspondence.

\begin{figure}[t]
  \centering
  \includegraphics[width=\columnwidth]{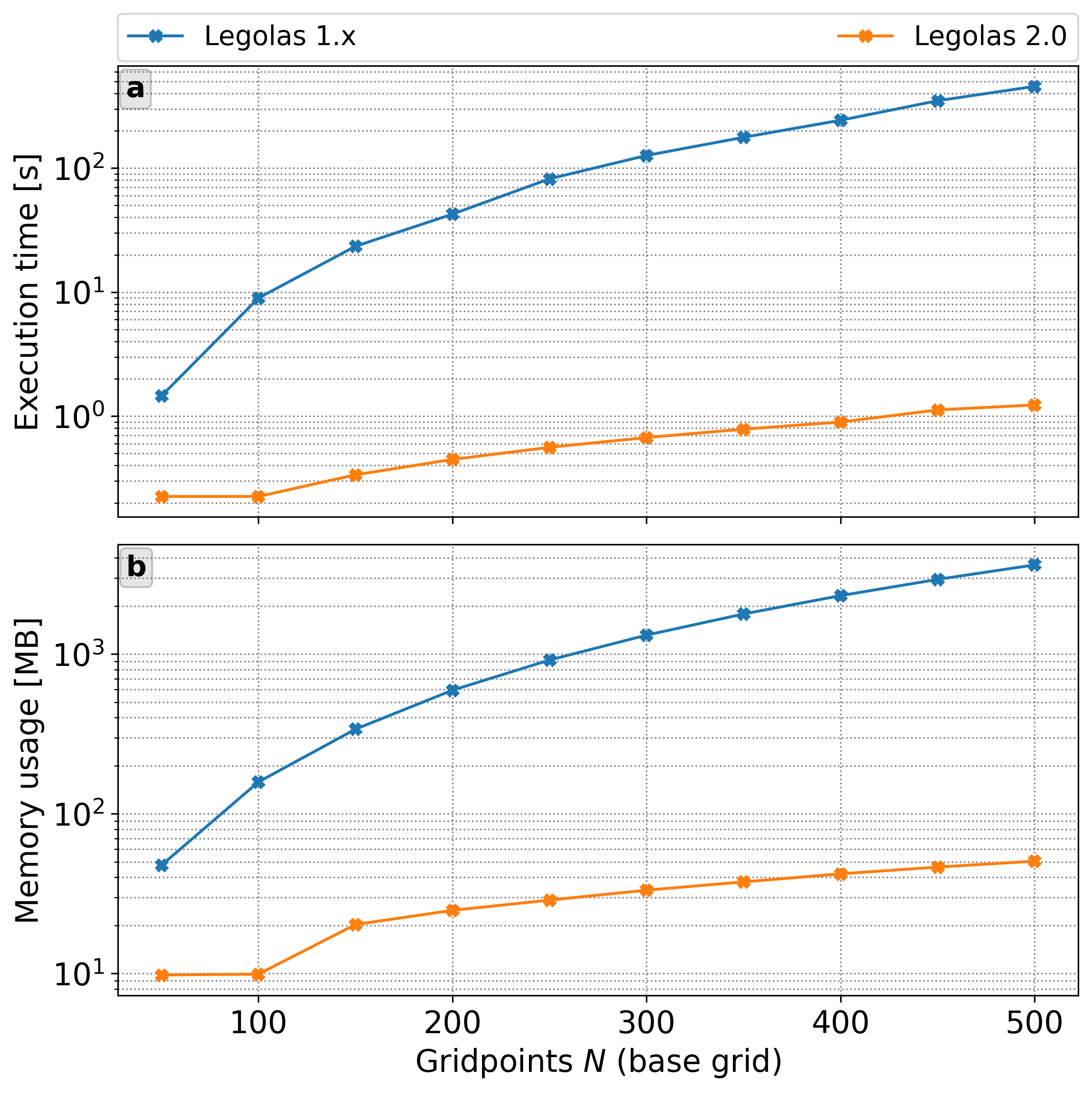}
  \caption{
    Execution time (Panel a) and memory usage (Panel b) for Arnoldi shift-invert. The blue and orange curves represent
    {\legolasOne} and {\legolasTwo}, respectively. Performance increases by a few orders of magnitude, especially at
    higher resolutions.
  }
  \label{fig: arnoldi_benchmarks}
\end{figure}

All runs performed in this subsection yield the 20 most global modes, which are already resolved at low resolutions. For example, if we consider 1000 gridpoints as the resolved baseline then the maximum absolute difference in eigenvalues at 100 gridpoints is approximately $10^{-5}$,
whereas for 500 gridpoints this decreases to roughly $10^{-9}$. Hence no extreme resolutions are needed here to properly resolve the sequence of unstable eigenmodes, and their corresponding eigenfunctions do not show strong variations as a function of radius such that lower resolutions are sufficient to resolve those as well. However, if a shift-invert run would be performed on the continued, stable part of the sequence, one would expect to encounter more fine-scaled eigenfunctions. Panel \textbf{b} in Figure \ref{fig: spectrum_si} shows the real part of the density eigenfunctions for the 19th and 20th mode in the sequence, which already oscillate stronger compared to lower mode numbers as in Figure \ref{fig: full_mri_spectrum}, panels \textbf{d}-\textbf{g}. If the 140th mode in the sequence were to be probed (which resides in the stable part), much higher resolutions would be needed to fully resolve its fine-scale eigenfunction structure.

\begin{figure}[t]
  \centering
  \includegraphics[width=\columnwidth]{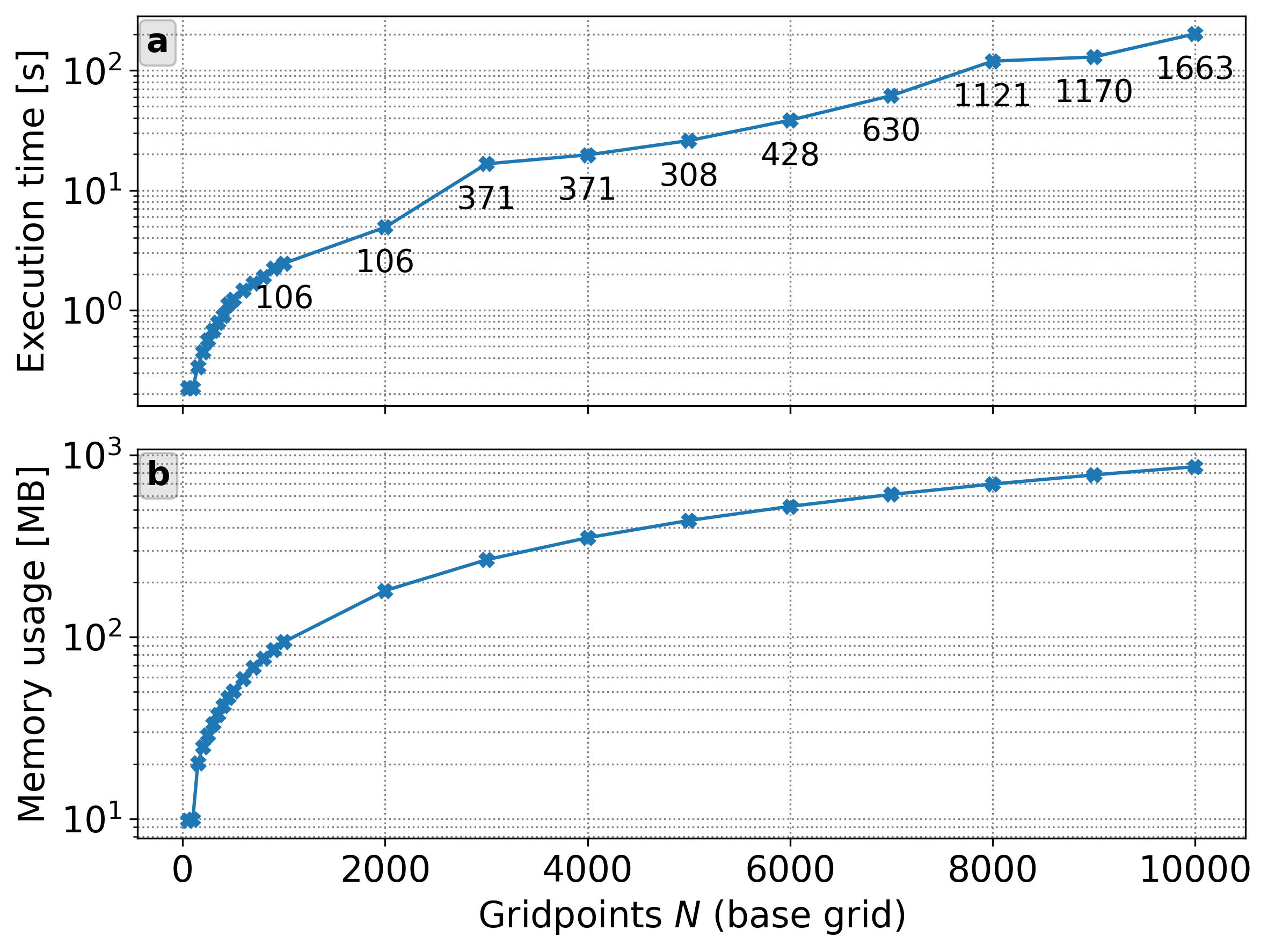}
  \caption{
    Execution time (Panel a) and memory usage (Panel b) for the {\legolasTwo} shift-invert solver at high grid resolutions. The numbers on Panel a indicate the amount of matrix-vector products done during the iteration as
    reported by ARPACK.
  }
  \label{fig: arnoldi_highres}
\end{figure}

Figure \ref{fig: arnoldi_benchmarks} shows the execution time (panel \textbf{a}) and maximum memory usage (panel \textbf{b}) for Arnoldi shift-invert. For 500 gridpoints the execution time decreases from $\approx 460$ seconds to approximately one second, with memory usage decreasing from $\approx 3.6$ GB to $50$ MB, corresponding to a performance increase of over two orders of magnitude. Note that the $y$-axis scale is logarithmic in both panels. Further scaling is shown in Figure \ref{fig: arnoldi_highres}, where the performance of the updated shift-invert solver is shown for rather extreme grid resolutions. Even for $10^4$ gridpoints, corresponding to matrix dimensions of $160000\times160000$, memory usage is still below 1 GB and convergence is achieved after only 200 seconds. The numbers underneath the datapoints in panel \textbf{a} represent the amount of matrix-vector products done during the iteration, as reported by ARPACK itself.

\subsection{Performance of inverse iteration} \noindent
To close this section we look at the performance of inverse vector iteration (IVI) as a function of gridpoints, taking a shift $\sigma = -0.003 + 0.6\itxt$.
For the convergence criterion \eqref{eq: inverse_criterion} we take $\epsilon = 10^{-9}$. It should be noted that if $\epsilon$ is taken to be too small, the combination of $\epsilon$ and the eigenvalue magnitude could cause the criterion to drop below machine precision, which typically means that convergence will never be achieved.

\begin{figure}[t]
  \centering
  \includegraphics[width=\columnwidth]{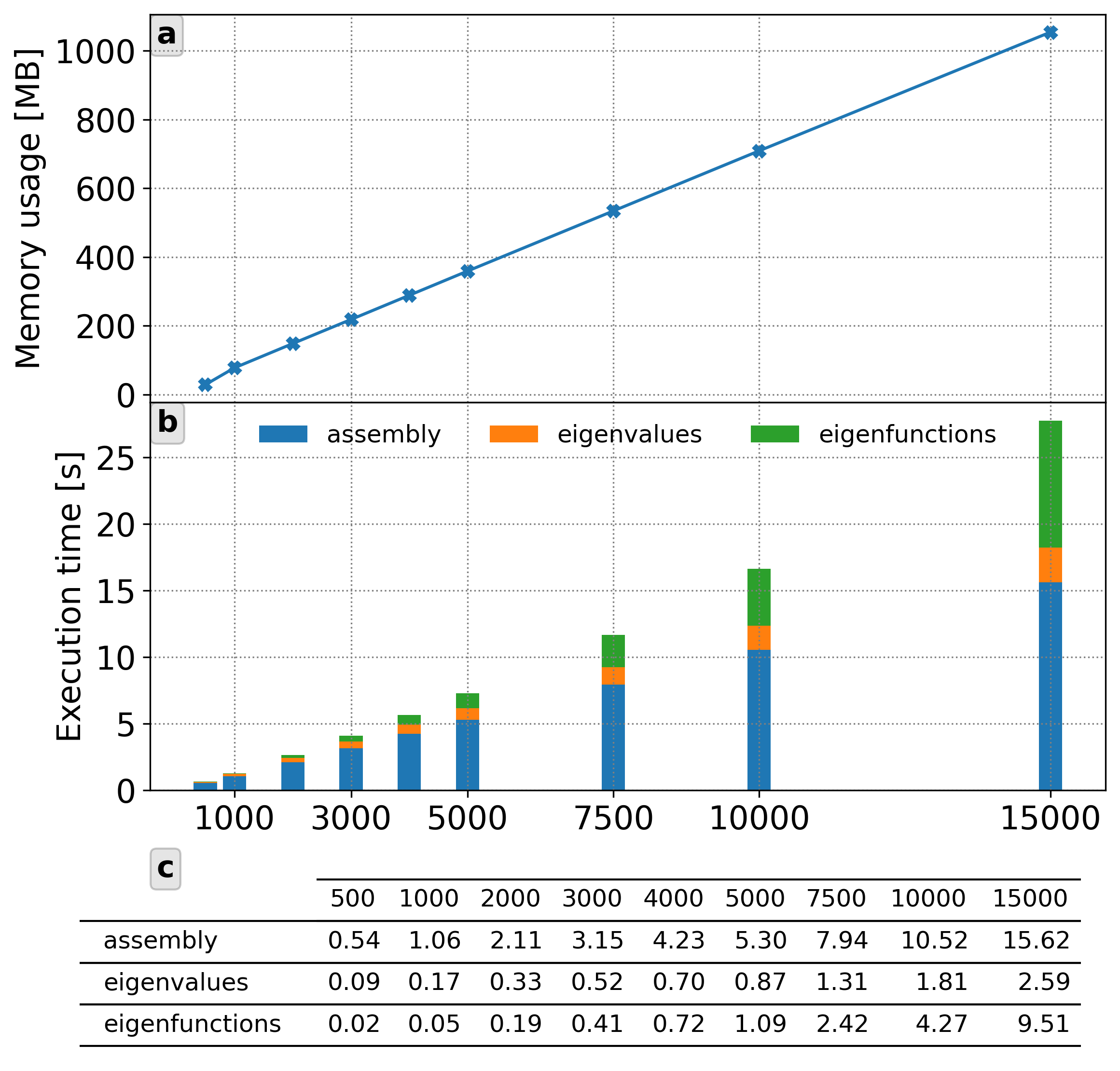}
  \caption{
    Memory usage (Panel a) and execution time (Panel b) as a function of resolution for inverse iteration with
    $\sigma = -0.003 + 0.65\itxt$ and $\epsilon = 10^{-9}$. Panel c shows the time needed in seconds for matrix assembly (blue), the actual iteration (orange) and assembly of the eigenfunctions (green).
  }
  \label{fig: inverse_iteration}
\end{figure}

Figure \ref{fig: inverse_iteration} shows memory usage (panel \textbf{a}) and execution time (panel \textbf{b}) for inverse iteration using the aforementioned shift and equilibrium background, in all cases convergence occurs in less than 10 iterations. Furthermore, the actual time needed for the iteration is only a fraction of the total runtime (see panel \textbf{b}). For example, running this setup at 15000 gridpoints takes roughly 28 seconds, of which 15.6 seconds are spent on assembly of the matrices ($\approx 56\%$), that is, calculating all matrix elements across the entire grid and approximating the integrals using Gaussian quadrature as explained in \cite{claes2020_legolas}. Assembly of the eigenfunctions takes approximately 9.5 seconds ($\approx 34\%$); the inverse iteration process itself only needs about 2.5 seconds ($\approx 9\%$) to converge. The remaining one percent is spent on initialisation, creation of the save file and additional overhead.

It should be noted that there is a tradeoff between shift-invert and inverse iteration, which depends on the number of modes that are requested and their location in the complex plane; if only a couple modes are needed then inverse iteration will usually be faster. For a larger number of modes (say, $\geq 15$) shift-invert is the better choice, especially if the modes are clustered since that will lead to faster convergence \cite{book_arpack}. For modes scattered in the complex plane shift-invert will take longer; however, if their locations are approximately known a priori a good starting guess can be provided to IVI. Generally speaking if $n$ modes are requested one should make the following consideration which case is more convenient and/or efficient: running IVI $n$ times, each time returning a single mode; or running inverse iteration one time for $n$ modes, returning them all at once.

Table \ref{tab: ivi_sequence} shows the first 10 unstable modes in the sequence calculated using IVI at 2000 gridpoints, where each shift was taken approximately equal to the results from shift-invert. The tolerance $\epsilon$ was taken to be $10^{-9}$ in all cases, convergence was achieved in under five iterations. All results match those from shift-invert up to absolute differences of approximately $10^{-8}$. Total runtime for all 10 runs is approximately 20 seconds, compared to 5 seconds for shift-invert at the same resolution (which is for 20 eigenvalues), see Figure \ref{fig: arnoldi_highres}.

\begin{table}[t]
  \centering
  \caption{
    Eigenvalues of the first 10 instabilities using IVI, using 2000 gridpoints and tolerance $\epsilon = 10^{-9}$.
  }
  \begin{tabular}{cc}
    \T
    Mode & IVI eigenvalue \\
    \hline
    \T
    $\omega_1$ & $-0.0020312088+0.6277216118\itxt$ \\
    $\omega_2$ & $-0.0018630009+0.5804872708\itxt$ \\
    $\omega_3$ & $-0.0017420272+0.5443808079\itxt$ \\
    $\omega_4$ & $-0.0016456679+0.5141323928\itxt$ \\
    $\omega_5$ & $-0.0015650883+0.4876853393\itxt$ \\
    $\omega_6$ & $-0.0014957010+0.4639637596\itxt$ \\
    $\omega_7$ & $-0.0014349990+0.4422720000\itxt$ \\
    $\omega_8$ & $-0.0013810000+0.4221620000\itxt$ \\
    $\omega_9$ & $-0.0013320000+0.4033330000\itxt$ \\
    $\omega_{10}$ & $-0.0012870000+0.3855560000\itxt$ \\
  \end{tabular}
  \label{tab: ivi_sequence}
\end{table}

\section{Subsystems} \label{sect: subsystems} \noindent
\begin{figure*}[t]
  \centering
  \includegraphics[width=\textwidth]{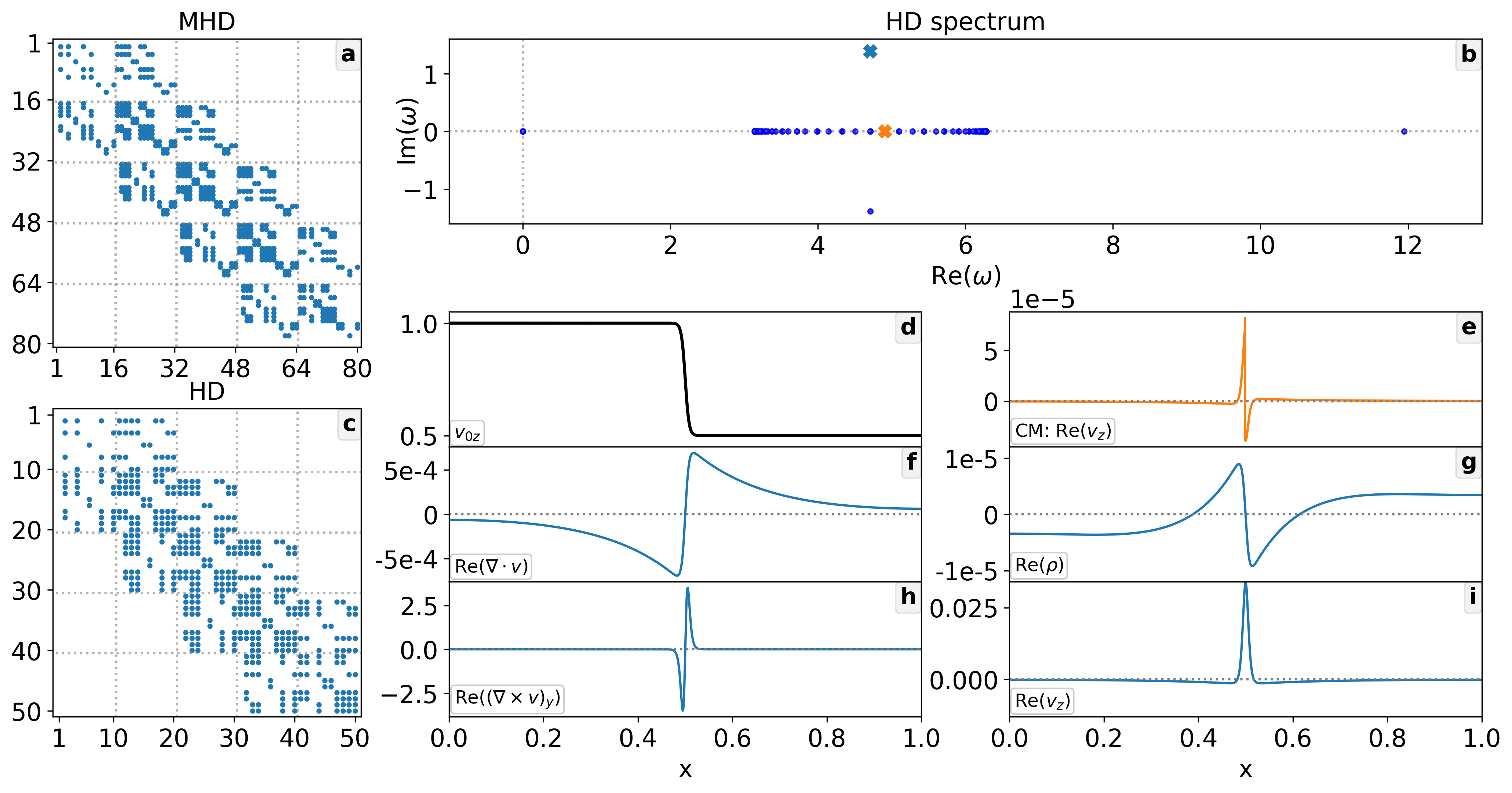}
  \caption{
    Spectra, matrices and eigenfunctions for the hydrodynamic setup discussed in Subsection \ref{sect: setup_khi}.
    Panels a and c: matrix for five gridpoints for {\legolas}' MHD and HD mode, respectively, indicating a significant
    reduction in size. Panel b: hydrodynamic spectrum calculated with QR-Cholesky at 250 gridpoints, blue and orange crosses annotate the KHI and a continuum mode, respectively. Panel d: the background velocity component $v_{0z}$. Panel e: $v_z$ eigenfunction for the continuum mode. Other panels: real part of the KHI eigenfunctions corresponding to the velocity divergence (f), density (g), $y$-component of the velocity curl (h), and $v_z$ (i).
  }
  \label{fig: subsystems}
\end{figure*}
Thus far, the linearised equations that were solved correspond to the 8-component vector given in \eqref{eq: state_vector}, that is, the full set of eight linearised MHD equations. In some cases we may have interest in pure hydrodynamic setups, or in cases with only 1D density/velocity/temperature variations, or want to anticipate special cases like incompressible flow conditions, or to use specific closure relations between density-pressure-temperature that omit the energy equation. One might also want to use to the framework to extend spectroscopy to multi-fluid PDE's. For these purposes we need to ensure that the {\legolas} framework can handle less (or more) than eight component entries in its state vector.

\subsection{Details} \noindent
Previously, {\legolasOne} was already able to handle pure hydrodynamic setups. However, in all cases the full set of linearised MHD equations was solved even if there was no magnetic field present. Hydrodynamics was handled by simply setting the magnetic field components to zero, resulting in a lot of zero rows and columns in the $\amat$ matrix. Furthermore, not all matrix elements associated with the magnetic vector potential components $(a_1, a_2, a_3)$ contain the magnetic field. Some of them also contain the wave numbers, background velocity field components or resistivity, without magnetic field components, such that simply setting $\bfb_0 = 0$ does not necessarily always result in $0 = 0$ rows. Hence, running hydrodynamic setups for an MHD state vector implies that a lot of unnecessary calculations are performed when solving the eigenvalue problem, resulting in zero eigenvalues that do not contribute to the spectrum; this follows from the fact that the HD and MHD spectra must be equal in the absence of a background magnetic field.

An additional feature of {\legolasTwo} is the possibility to run subsystems of the set of MHD equations, made possible by the dynamic nature of matrix assembly through the linked-list implementation. More specifically, if hydrodynamics is specified, the state vector given in \eqref{eq: state_vector} reduces to
\begin{equation} \label{eq: state_vector_hd}
  \bfx = \left(\rho_1, v_1, v_2, v_3, T_1\right)^\top.
\end{equation}
When assembling the matrices every matrix element undergoes a check whether their associated state vector components are present in the state vector \eqref{eq: state_vector_hd}. If an element does not appear it is skipped and never added to the matrix, and because the assembly process is dynamic there is no (re)allocation of the matrices needed.
Similar reasonings can be made for state vectors with only part of the 3D velocity/magnetic field vector components, or without an energy equation.

This feature implies a huge degree of freedom, both in terms of implementation and possible systems that can be investigated. First of all this allows us to run {\legolas} in 1D, 2D or 3D (magneto)hydrodynamics using the exact same implementation by modifying the associated state vector. For example, by taking a state vector $(\rho_1, v_1, T_1)^\top$ one can represent a one-dimensional plasma with a fixed field line shape. No cumbersome switches inside the source code are needed to deal with the various possible combinations, as elements can simply be added, skipped, or removed at will.

Second, by extending the state vector with additional components other systems can be modelled as well. Examples include supporting self-gravity (that is, adding the Poisson equation), multi-fluid (M)HD and even optically thick radiative treatments. These additions require a detailed calculation of new matrix elements and their implementation, though they can be readily tied in to the existing framework as the matrices will dynamically adapt to their new sizes.

\subsection{Example setup: KHI} \label{sect: setup_khi} \noindent
As an example we consider a relatively simple setup, that is, a homogeneous background given by $\rho_0 = 1$ and
$T_0 = 1$ in normalised units. The geometry is Cartesian with $x \in [0, 1]$ and we consider hydrodynamics, so there is no background magnetic field present. A hyperbolic tangent profile is imposed for the $z$-component of the background velocity field with a sharp transition layer in the center of the domain such that we have two counterstreaming flows, implying that the setup is prone to Kelvin-Helmholtz instabilities (KHI). The $v_{0z}$ component is given by
\begin{equation} \label{eq: tanh_profile}
  v_{0z}(x) = \frac{\beta C(x_0) - \alpha C(x_1) + (\alpha - \beta)C(x)}{C(x_0) - C(x_1)},
\end{equation}
in which the function $C(u)$ is given by
\begin{equation}
  C(u) = \tanh\left(\frac{2\pi(s - u)}{\delta}\right).
\end{equation}
The points $x_0 = 0$ and $x_1 = 1$ correspond to the left and right edges of the domain, respectively; $s = 0.5$ denotes the centre of the profile and $\delta = 0.05$ its width. The variables $\alpha = 1.0$ and $\beta = 0.5$ are constants indicating the background velocity values in $x_0$ and $x_1$, respectively, taken as such to break left-right symmetry in the spectrum. This particular velocity profile also ensures a sharp but smooth velocity transition without discontinuities. The wave numbers are taken to be $k_y = 0$ and $k_z = 2\pi / (x_1 - x_0)$ in order to have a wave vector component along the direction of flow.

The matrices for this particular setup are shown in Figure \ref{fig: subsystems} for five gridpoints $N$, with {\legolas} running in MHD mode (panel \textbf{a}) and HD mode (panel \textbf{c}); the matrices differ in size considerably. For MHD, associated with the state vector \eqref{eq: state_vector}, the matrices have dimensions $80\times80$ (i.e. $16N$); while for HD, associated with the state vector \eqref{eq: state_vector_hd}, the matrices have dimensions $50\times50$ (i.e. $10N$). It is clear that for large resolutions this difference has a major impact on performance.
The hydrodynamic spectrum for this setup, calculated using QR-Cholesky at 250 gridpoints, is shown in Figure \ref{fig: subsystems}, panel \textbf{b}, and is the same as the MHD spectrum. For reference, the MHD calculation took roughly 1 minute to complete, while the exact same calculation using the HD subsystem finished in 19 seconds. As expected there is an instability present, annotated with cross, corresponding to the KHI. A complex conjugate (stable) mode is also present due to up-down spectral symmetry. The dense collection of eigenvalues on the real axis below the instability correspond to the Eulerian (entropy) continuum. Panel \textbf{d} shows the background velocity profile as given by Equation \eqref{eq: tanh_profile}. Panel \textbf{e} shows the real part of the $v_z$ eigenfunction for a mode in the continuum, annotated with an orange cross. The other panels contain the real part of some eigenfunctions for the KHI mode (blue cross), showing the velocity divergence (f), density (g), $y$-component of the velocity curl (h), and $v_z$ (i). All eigenfunctions were calculated using IVI at 2000 gridpoints.

\begin{figure*}[t]
  \centering
  \includegraphics[width=\textwidth]{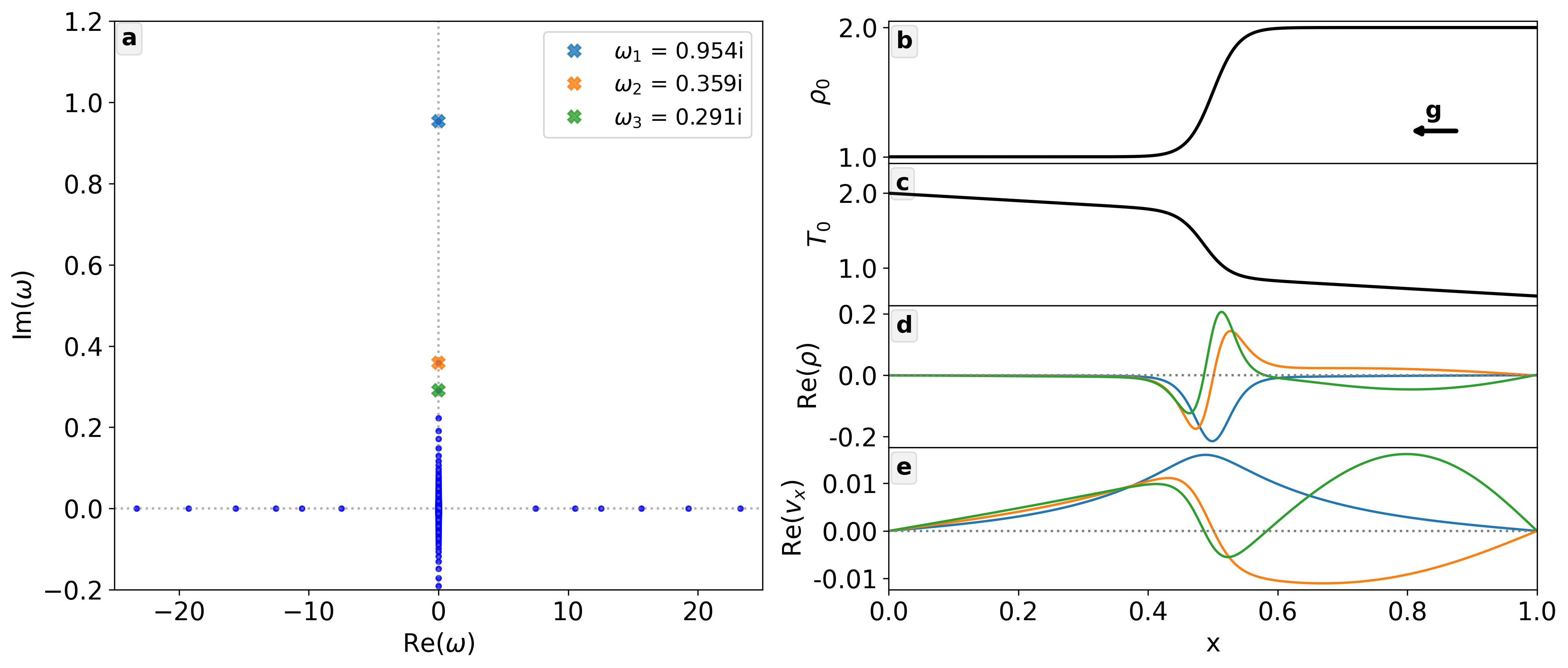}
  \caption{
    Spectrum (a), background profiles (b-c) and eigenfunctions (d-e) of the annotated modes, corresponding to the
    Rayleigh-Taylor setup described in Section \ref{sect: setup_rti}. The direction of gravity is annotated with an arrow.
  }
  \label{fig: RTI}
\end{figure*}

\subsection{Example setup: RTI} \label{sect: setup_rti} \noindent
As a second example we consider Rayleigh-Taylor instabilities where the background density $\rho_0(x)$ uses the same prescription as Equation \eqref{eq: tanh_profile}. The background temperature $T_0$ can then be obtained by integrating the force-balance equation $\left(\rho_0 T_0\right)' + \rho_0 g = 0$ where the constant $g$ represents the uniform gravitational field downwards along $x$ and the prime denotes the derivative with respect to $x$. Taking the integration constant such that $T_\text{c}$ represents the (dimensionless) temperature at $x = x_0$, this yields

\begin{equation} \label{eq: T0_RTI}
  \begin{gathered}
    T_0(x) = T_\text{c} \\
    + \frac{
      g\delta(\alpha - \beta)\Bigl(\ln\bigl(C(x_0) + 1\bigr) - \ln\bigl(C(x) + 1\bigr)\Bigr)
    }{2\pi \rho_0(x)\bigl(C(x_0) - C(x_1)\bigr)} \\
    + \frac{
      g(x - x_0)\bigl(\alpha C(x_1) - \alpha - \beta C(x_0) + \beta\bigr)
    }{\rho_0(x)\bigl(C(x_0) - C(x_1)\bigr)}.
  \end{gathered}
\end{equation}
Here $\alpha$ and $\beta$ are taken to be 1 and 2, respectively, representing a heavier fluid on top of a lighter fluid (Atwood number of 1/3). The profile's width equals $\delta = 0.25$ and is centered at $s = 0.5$, the geometry is Cartesian with $x \in [0, 1]$, $g = 0.5$ and $T_\text{c}$ equals 2. The wave numbers $k_y$ and $k_z$ are taken to be 0 and $2\pi/(x_1 - x_0)$, respectively.

Figure \ref{fig: RTI} shows the spectrum associated with this particular setup in panel \textbf{a}, revealing a sequence of purely imaginary eigenmodes corresponding to the expected Rayleigh-Taylor instabilities. The first three modes in this sequence have been annotated with crosses. Panels \textbf{b} -- \textbf{c} show the background density and temperature profiles, respectively, corresponding to Equations \eqref{eq: tanh_profile} and \eqref{eq: T0_RTI}. The arrow on panel \textbf{b} indicates the direction of the (constant) gravitational field. Panels \textbf{d} -- \textbf{e} show the real part of the density and $v_x$ eigenfunctions associated with the three annotated modes, which will be used for visualisation purposes in Section \ref{sect: visualisations}. The spectrum and eigenfunctions were calculated using QR-Cholesky at 250 gridpoints, with {\legolas} running in HD mode similar to the KHI case earlier.

\section{Eigenfunction visualisations} \label{sect: visualisations} \noindent
The final new feature in {\legolasTwo} is related to its post-processing framework {\pylbo}, and introduces the possibility to visualise eigenfunctions in multiple dimensions, thereby relying on the wave mode representation as given by Equation \eqref{eq: fourier}. {\legolas} calculates the eigenfrequencies $\omega$ and their associated eigenfunctions $\hat{f}_1(u_1)$ for given wave numbers $k_2$ and $k_3$, hence the temporal and spatial coordinates
$(t, u_2, u_3$) can be chosen at will to generate various datacubes. More specifically, we can now visualise eigenfunctions in the following ways, for both Cartesian and cylindrical geometries:
\begin{enumerate}
  \item \textit{1D temporal}. Fixed values for $u_2$, $u_3$ and a range in $t$. This generates a time-position plot showing the temporal evolution of an eigenfunction on a 2D colormap.
  \item \textit{2D spatial along $u_3$}. Fixed values for $u_3$, $t$ and a range in $u_2$. This can be thought of as a slice through a 3D domain at a specific $u_3$ value. For Cartesian geometries this results in a $xy$-plot (and will hence be circular for cylindrical geometries).
  \item \textit{2D spatial along $u_2$}. Fixed values for $u_2$, $t$ and a range in $u_3$, i.e. a slice for a fixed $u_2$ coordinate. For Cartesian geometries this results in a $xz$-plot; for cylindrical geometries this implies slicing at a fixed $\theta$-value, thus resulting in a rectangular $zr$-plot.
  \item \textit{3D spatial}. The same as for the 2D spatial case but now for a range in $u_2$ or $u_3$ values.
  The 2D slices will be stacked to produce a 3D view.
  \item \textit{VTK export}. Exporting a full 3D datacube to \textsf{.vtk} format is also supported, where both the eigenfunctions and equilibrium background can be saved to the same file. This is compatible with advanced visualisation tools such as \textsf{ParaView} and/or \textsf{VisIt}, which can be used to trace for example magnetic or velocity field lines for both the equilibrium background and the perturbed quantities.
\end{enumerate}
Furthermore, all visualisations can be evolved in time. This allows the user to generate movies where every snapshot will be a 2D/3D view at a particular timestep $t_i$. Additionally the eigenfunctions of multiple eigenvalues can be superimposed, where Equation \eqref{eq: fourier} is calculated independently for each one and then summed.
In all visualisations either the real or imaginary part of the final result can be taken. Note that we can always multiply the eigenfunctions with an arbitrary complex factor, which is nothing more than applying a phase shift and/or an amplitude modulation. In what follows the Kelvin-Helmholtz mode refers to the annotated mode on Figure \ref{fig: subsystems}, while the Rayleigh-Taylor modes refer to the three annotated modes on Figure \ref{fig: RTI}.

\subsection{1D temporal view} \label{sect: 1d_temporal}
\begin{figure}[t]
  \centering
  \includegraphics[width=\columnwidth]{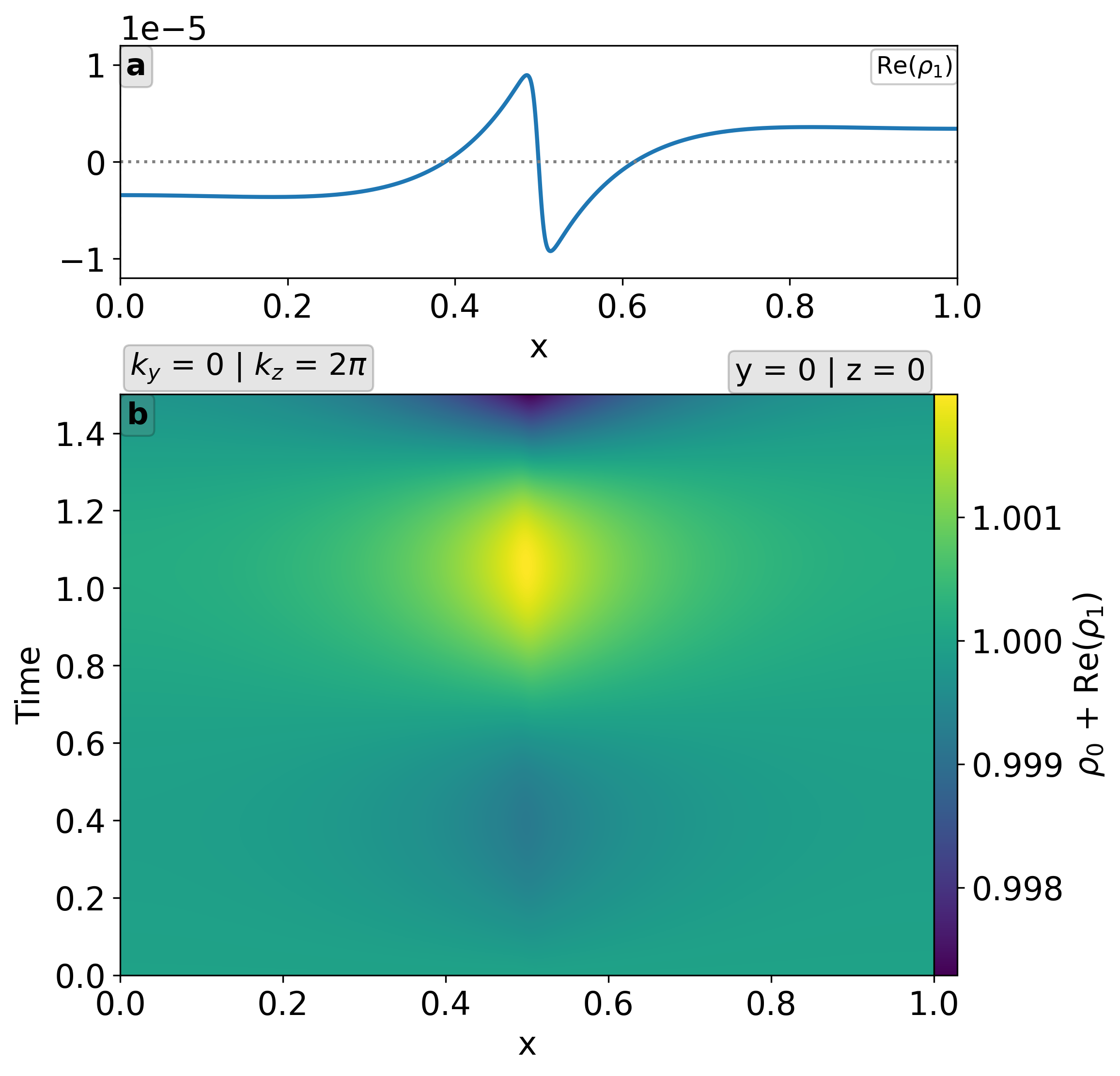}
  \caption{
    Panel b: temporal evolution of the density eigenfunction for the KHI as given in Figure \ref{fig: subsystems}, the (uniform) background is superimposed. The eigenfunction itself is shown on Panel a.
  }
  \label{fig: 1d_temporal}
\end{figure}

\noindent
Figure \ref{fig: 1d_temporal}, panel \textbf{b}, shows a one-dimensional temporal view of the density eigenfunction associated with the Kelvin-Helmholtz mode, superimposed on the (constant) density background. For clarity the eigenfunction is shown again in panel \textbf{a} as a function of position. Since both $y$ and $z$ are taken to be zero the pure temporal dependence is shown here, with density values alternating near the interface.

\subsection{2D spatial view}
\begin{figure}[t]
  \centering
  \includegraphics[width=\columnwidth]{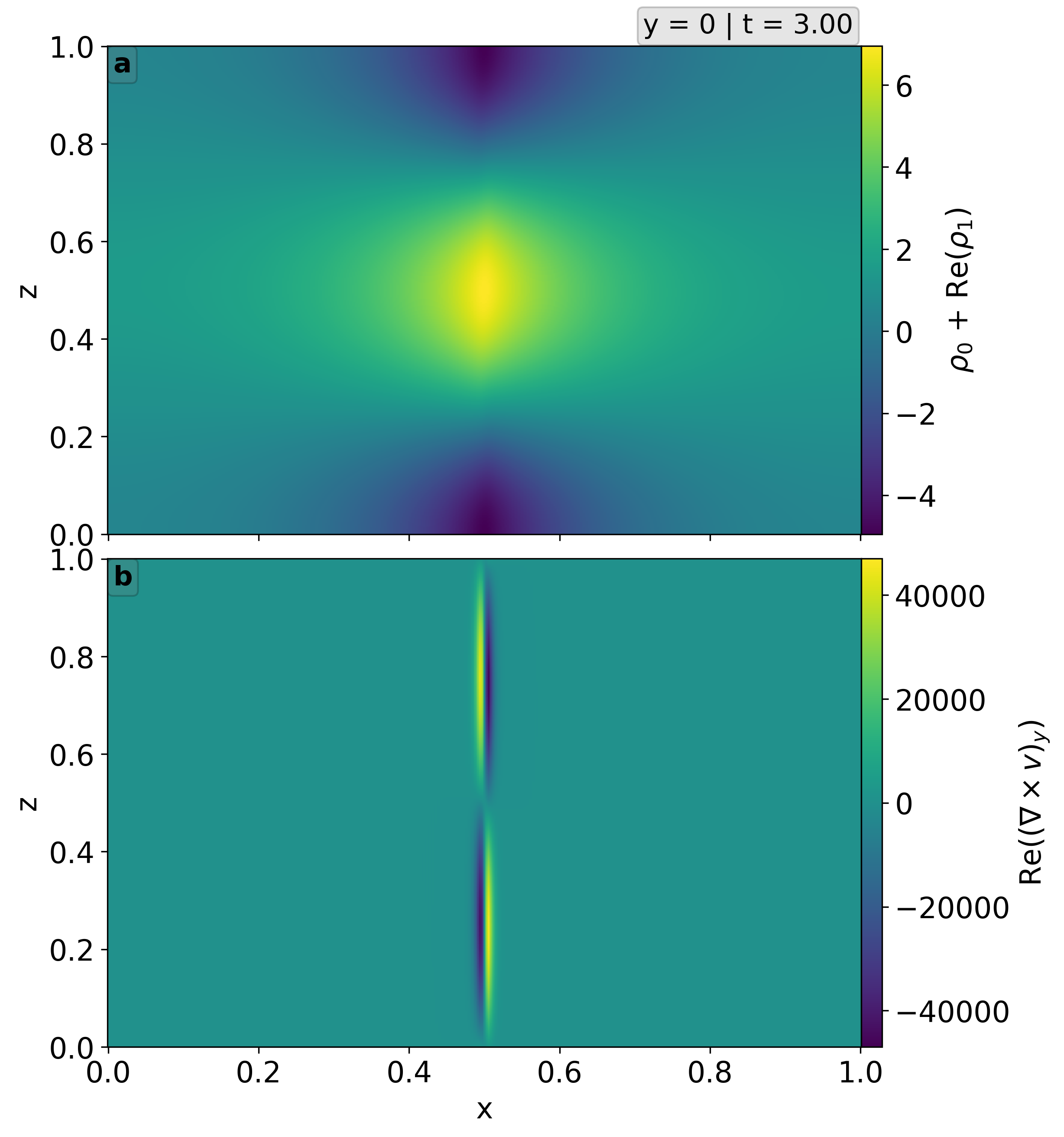}
  \caption{
    2D spatial view along $y = 0$ for the density eigenfunction (Panel a) and $y$-component of the vorticity (Panel b), for the KHI eigenfunction at $t = 3$.
  }
  \label{fig: 2d_spatial}
\end{figure}

\noindent
A two-dimensional view taken along $y = 0$ is given in Figure \ref{fig: 2d_spatial}, showing the density perturbation and background (panel \textbf{a}) along with the $y$-component of the velocity curl. The range in $z$ can be chosen arbitrarily and is taken as $z \in [0, 1]$, along with $t = 3$. Vorticity is, as expected, highly localised near the density interface. As $x$ denotes height, the alternating magnitude of the vorticity as a function of $z$ may give rise to the typical vortices one expects for Kelvin-Helmholtz instabilities. It should be noted that these views purely represent the linear stage and hence cannot show nonlinear phenomena.

\subsection{3D spatial view}
\begin{figure}[t]
  \centering
  \includegraphics[width=\columnwidth]{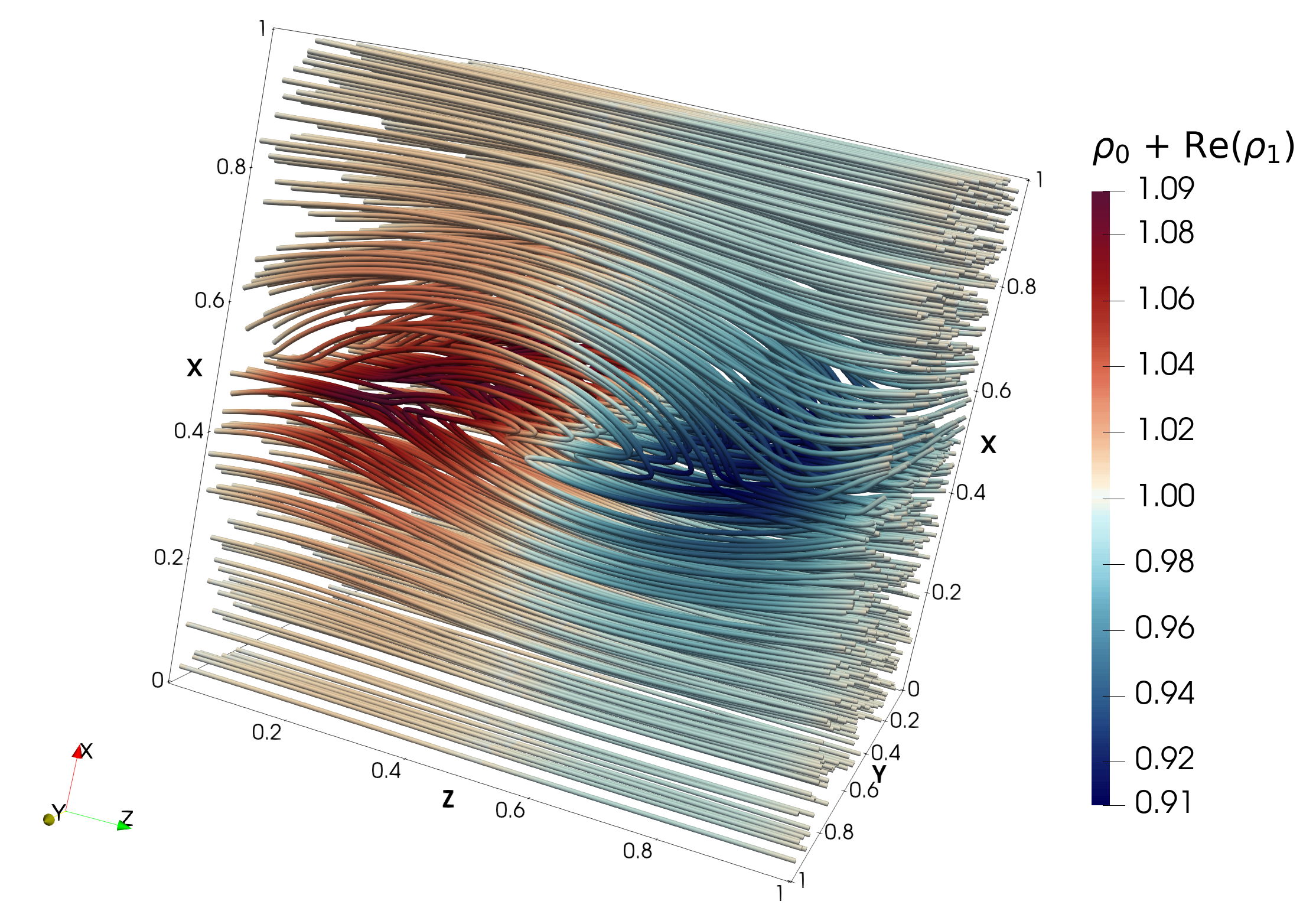}
  \caption{
    3D streamline visualisation for the KHI at $t = 0$, showing the total velocity field coloured by total density, showing clear indications of vortex formation already in the early linear stage. An animation of this figure is available online.
  }
  \label{fig: 3d_khi_vfield}
\end{figure}
\begin{figure}[t!]
  \centering
  \includegraphics[width=\columnwidth]{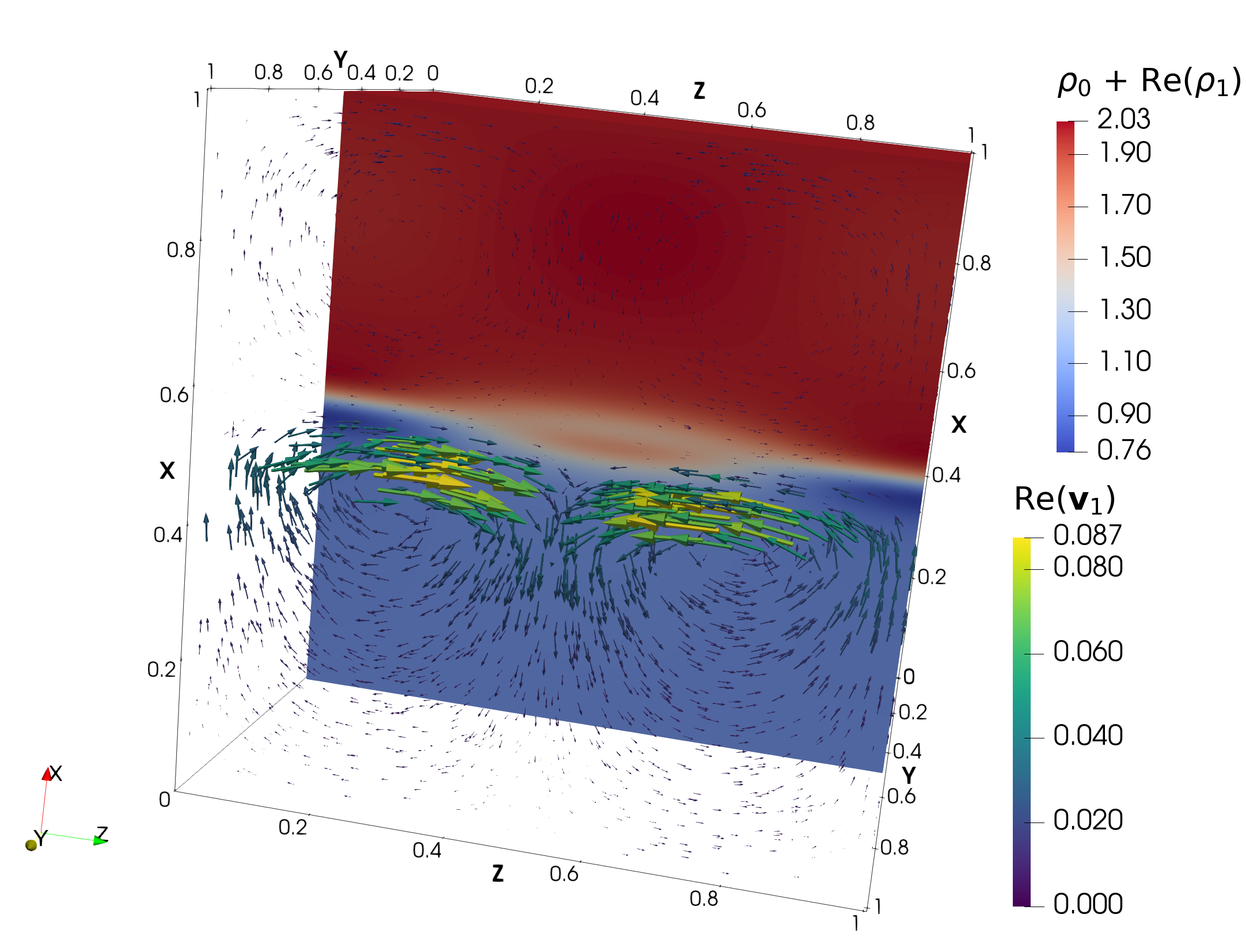}
  \caption{
    3D visualisation for a superposition of the three RTI modes at $t = 0$, arrows indicate the velocity field and are coloured and scaled to its magnitude, total density is also shown.
  }
  \label{fig: 3d_rti_vfield}
\end{figure}

\noindent
As a final application we show 3D visualisations obtained by exporting the eigenmode solutions and equilibrium backgrounds to \textsf{.vtk} format. The perturbed velocity fields, for example, can then be constructed through $v_1\unit{x} + v_2\unit{y} + v_3\unit{z}$, where $(v_1, v_2, v_3)$ denote the eigenfunctions and $(\unit{x}, \unit{y}, \unit{z})$ the triad of unit vectors; the background velocity field can be assembled in a similar way.

Figure \ref{fig: 3d_khi_vfield} shows the total perturbed velocity field superimposed on the background flow for the Kelvin-Helmholtz mode at $t = 0$, the real part of the eigenmode solutions was always taken. Velocity streamlines are coloured to total density as indicated by the colourbar. As can be seen on Figure \ref{fig: subsystems} the velocity and density eigenfunctions are rather localised near the interface $x = 0.5$, such that near the edges of the domain the background flow (purely directed along $\unit{z}$) dominates. Closer to the interface the streamlines reveal more intricate behaviour as the combined spatial dependence of the $v_x$ and $v_z$ eigenfunctions comes into play. The resulting structure resembles the initial stages of vortex formation near the interface, as expected for Kelvin-Helmholtz instabilities. Note that $k_y = 0$, such that the structure propagates in the $z$-direction. An animation of this figure is available online. Furthermore these kind of views purely represent the linear stage and as such cannot show nonlinear phenomena. In actual simulations however one could expect that the early stages of the simulation closely resemble this view, after which nonlinear effects will take over followed by the emergence of ``true'' KHI vortices.

Figure \ref{fig: 3d_rti_vfield} shows a superposition of the three most unstable RTI modes at $t = 0$: the eigenmode solution for each individual mode is calculated, these are summed, and the real part of the result is taken. Note that here $k_y = 0$ such that the solutions along $\unit{y}$ are invariant. Total density is shown in the background with the localised density eigenfunctions resulting in a slightly higher density region near the interface. The velocity field is shown with arrows and are coloured and scaled to the field magnitude, note that gravity points downwards along $\unit{x}$. The resulting view is typical for a Rayleigh-Taylor instability: a density perturbation near the interface accompanied by two counter-rotating vortices, the resulting torque on the interface will cause a downwards displacement of the perturbation, eventually resulting in the typical finger-like structures associated with the RTI.

\section{Conclusions} \label{sect: conclusions} \noindent
In this paper we report on various improvements and extensions to the open-source {\legolas} code, suited for MHD spectroscopy of general three-dimensional equilibria with a nontrivial one-dimensional variation. A new linked-list based datastructure was implemented to store the matrices in the eigenvalue problem, resulting in a considerable reduction of memory usage. Additionally, matrix-vector products are now more efficient, which significantly speeds up computational time when solving the eigenvalue problem compared to the dense treatment of the previous {\legolasOne} version.

The existing QR-based solver (which previously relied on an explicit inversion of the $\bmat$-matrix) has been updated to use a LU-decomposition of $\bmat$ instead, along with a linear system solver, to reduce the general eigenvalue problem to standard form without explicit inversion. A second solver has been implemented in case the $\bmat$-matrix is symmetric, which relies on the Cholesky factorisation of $\bmat$ instead of a LU-decomposition, and should be more efficient and numerically stable. Various benchmarks were performed, showing that QR-LU and QR-Cholesky are more than $50\%$ more efficient in terms of execution time and memory usage compared to QR-invert from {\legolasOne}.

The previous Arnoldi-based solvers have also been updated, that is, both the general solver and shift-invert spectral transformation, and now efficiently exploit matrix sparsity in the eigenvalue problem. During benchmarking these updated routines showed an improvement of over two orders of magnitude, allowing users to perform extreme resolution runs without the need for HPC resources. A new type of solver was implemented, inverse vector iteration, which takes a user-defined shift in the complex spectrum and iterates towards the closest eigenvalue with corresponding eigenvector. The iterative process makes use of the Rayleigh quotient procedure, which increases the accuracy of the estimate in each iteration. Tests and benchmarks have shown that if a ``good'' guess is provided for the eigenvalue, convergence is achieved after only a few iterations, with performance slightly better than Arnoldi's shift-invert.

The framework now also includes the capability of running subsystems of the original set of eight linearised MHD equations, allowing for pure hydrodynamic setups or lower-dimensional cases. The reduced state vector results in a considerable decrease in matrix sizes, and thus the eigenvalue problem can be efficiently solved without a large number of zero rows or columns. The large degree of freedom associated with this functionality further increases the already wide application range of the framework, and also sets the stage for extensions with additional equations in the future.

Lastly we showcased the new visualisation feature of the code, which adds the possibility to visualise one or more mode eigenfunctions in multiple dimensions. This includes a one-dimensional temporal evolution, 2D/3D spatial views, and the capability to export eigenfunctions to \textsf{.vtk} format for advanced visualisations with external tools. The latter functionality was showcased on the well-known Kelvin-Helmholtz and Rayleigh-Taylor instabilities; their spectra were calculated and eigenfunctions of the unstable mode(s) exported to \textsf{.vtk}, and used to visualise density and velocity streamlines. Interestingly, the linear stages already showed strong indications of vortex formation for the KHI and typical counterstreaming flows near a density perturbation for the RTI, hinting at possible links between linear stability analysis and non-linear features.

Based on these results early conclusions can be drawn that eigenfunctions of the unstable mode(s) of a given system already provide information on general system behaviour and the onset of non-linear features, and is something that can be verified by performing true non-linear simulations of the same setup.
\newpage
\noindent
\small{
  \textit{Acknowledgements.}  The authors would like to thank Evert Provoost for useful contributions to the code, and Jordi De Jonghe for fruitful discussions. This work is supported by funding from the European Research Council (ERC) under the European Unions Horizon 2020 research and innovation programme, Grant agreement No. 833251 PROMINENT ERC-ADG 2018; and by internal funds KU Leuven, project C14/19/089 TRACESpace.
}

\bibliographystyle{elsarticle-harv}
\bibliography{bibfile.bib}

\end{document}